# Ultrafast imaging of polariton propagation and interactions

Ding Xu[1†], Arkajit Mandal[1†], James M. Baxter[1], Shan-Wen Cheng[1], Inki Lee[1], Haowen Su[1], Song Liu[2], David R. Reichman[1*], Milan Delor[1*]
1. Department of Chemistry, Columbia University, New York, NY 10027, United States
2. Department of Mechanical Engineering, Columbia University, New York, NY 10027, United States

[†] These authors contributed equally

**Abstract**
Semiconductor excitations can hybridize with cavity photons to form exciton-polaritons (EPs) with remarkable properties, including light-like energy flow combined with matter-like interactions. To fully harness these properties, EPs must retain ballistic, coherent transport despite matter-mediated interactions with lattice phonons. Here we develop a nonlinear momentum-resolved optical approach that directly images EPs in real space on femtosecond scales in a range of polaritonic architectures. We focus our analysis on EP propagation in layered halide perovskite microcavities. We reveal that EP–phonon interactions lead to a large renormalization of EP velocities at high excitonic fractions at room temperature. Despite these strong EP–phonon interactions, ballistic transport is maintained for up to half-exciton EPs, in agreement with quantum simulations of dynamic disorder shielding through light-matter hybridization. Above 50% excitonic character, rapid decoherence leads to diffusive transport. Our work provides a general framework to precisely balance EP coherence, velocity, and nonlinear interactions.

## Introduction

Exciton-polaritons (EPs) form when semiconductor excitons (electron-hole pairs) hybridize with photons, resulting in renormalization of light-matter eigenstates[1,2]. Hybridization is readily attained with or without photonic cavities at room temperature in two-dimensional (2D), hybrid and molecular semiconductors that sustain strong light-matter interactions and large exciton binding energies[3–6], paving the way to scalable polaritonic devices. EPs exhibit highly desirable properties. Photon-like EPs are highly coherent and exhibit long-range ballistic energy flow ideal for energy technologies[7–10]. Exciton-like EPs sustain ultra-strong nonlinear interactions that could lead to single-photon quantum switches[11–14]. Nevertheless, the true promise of EPs emerges when their light-like and matter-like features are combined. In this regime, many open questions remain: can EPs with high exciton character (>50%) preserve their intrinsic group velocities even in the presence of exciton-mediated scattering with phonons and other species? At what exciton fraction do matter-like interactions lead to EP decoherence? Maintaining long-range ballistic transport for highly matter-like excitations would enable, for example, large-scale photonic circuits based on single-photon quantum gates. Simultaneously optimizing EP transport and nonlinearities for any given system requires new high-throughput approaches capable of directly tracking EP propagation and interactions throughout their lifetimes.

Fast propagation and nonlinear interactions make EPs exceptionally challenging to study on their intrinsic spatiotemporal scales. Several powerful approaches have been developed to image EPs, though the majority rely on steady-state far-field optical microscopies[10,15–17] that are unable to directly track EP propagation and nonequilibrium processes. Some ultrafast implementations of nonlinear far-field microscopies have been applied to polaritonic systems[18–20], but importantly lack the momentum resolution or specie-specificity to probe EPs and their interactions with other material excitations. Other powerful implementations based on detecting polariton or condensate emission in the far field using streak cameras[21,22] or interferometry[23–25] enable real-space monitoring of polariton evolution, but require slow-moving, long-lived and brightly emissive species as well as high excitation fluences, limiting the generalizability of these approaches and the achievable signal-to-noise ratios in delicate measurements. Recent tour-de-force experiments using ultrafast near-field scanning[26–28] and electron microscopies[29] have been applied to non-cavity EPs and phonon-polaritons in van der Waals materials, boasting high momentum and spatial resolution, allowing direct tracking of polariton wavepackets on sub-picosecond scales. Nevertheless, near-field approaches are not generalizable to microcavity EPs and other complex material architectures that require sub-surface penetration, and the extension of electron microscopies to microcavity EPs and fragile materials remains untested.

Here we develop a highly generalizable and noninvasive approach based on spatiotemporally resolved far-field optical microscopy[30,31] allowing direct imaging of EP propagation and interactions on femtosecond–nanosecond scales with sub-100 nm spatial sensitivity. We term our approach Momentum-resolved Ultrafast Polariton Imaging (MUPI). By directly tracking EPs and excitons throughout their lifetimes with high spatiotemporal precision, we quantify key factors affecting EP propagation, including EP–lattice and EP–EP scattering as a function of light *versus* matter composition, quantities that were never directly empirically accessed. We demonstrate that our approach is generalizable across microcavity EPs, self-hybridized EPs in material slabs without external cavities, and plasmon-exciton (plexciton) polaritons in a range of emerging molecular and material systems. We focus our analysis on 2D halide perovskite microcavities at room temperature, an ideal test system that reaches the strong coupling limit without complicated cavity fabrication, possesses strong EP interactions[32], and displays low intrinsic exciton diffusivity[33,34] allowing unambiguous spatial isolation of excitonic versus EP signals. We reveal that EPs with large exciton character in these systems are strongly affected by scattering with lattice phonons, leading to up to 40% renormalization of EP velocity at room temperature. Remarkably, despite these strong EP–phonon interactions, EPs retain ballistic transport properties for up to 50% exciton character. Beyond 50% exciton character, matter-mediated interactions lead to decoherence and diffusive transport in our structures.

**Results**
**MUPI imaging of diffusive exciton and ballistic EP propagation**. MUPI is illustrated in Figure 1a (see Figures S1-S2 for more detail). A diffraction-limited femtosecond visible pump pulse generates excitons or EPs by exciting the material either above-gap or at a polariton resonance. A widefield backscattering probe[35] then images the sample with and without the pump pulse at controlled time delays. Differential pump ON/pump OFF images provide a direct readout of the spatial distribution of pump-generated species, which can be tracked with sub-diffraction spatial precision[30,31] (Supplementary Note 1). Momentum-matched probing of different EP modes (Figure 1b) in MUPI is achieved by displacing the probe along the optical axis of the objective, resulting in tilted widefield illumination of the sample[36]. The use of a high numerical aperture microscope objective (NA = 1.4) provides access to EP momenta $k = 2\pi\, \mathrm{NA}/\lambda$ greater than 10 μm$^{-1}$, where $\lambda$ is the wavelength of light. The sample can be imaged either in real space to track exciton and EP propagation (Figures 1c-g), or in momentum-space to provide the linear and nonlinear (excited state) EP dispersion through angle-resolved reflectance spectra (Figures 1b,h,i).

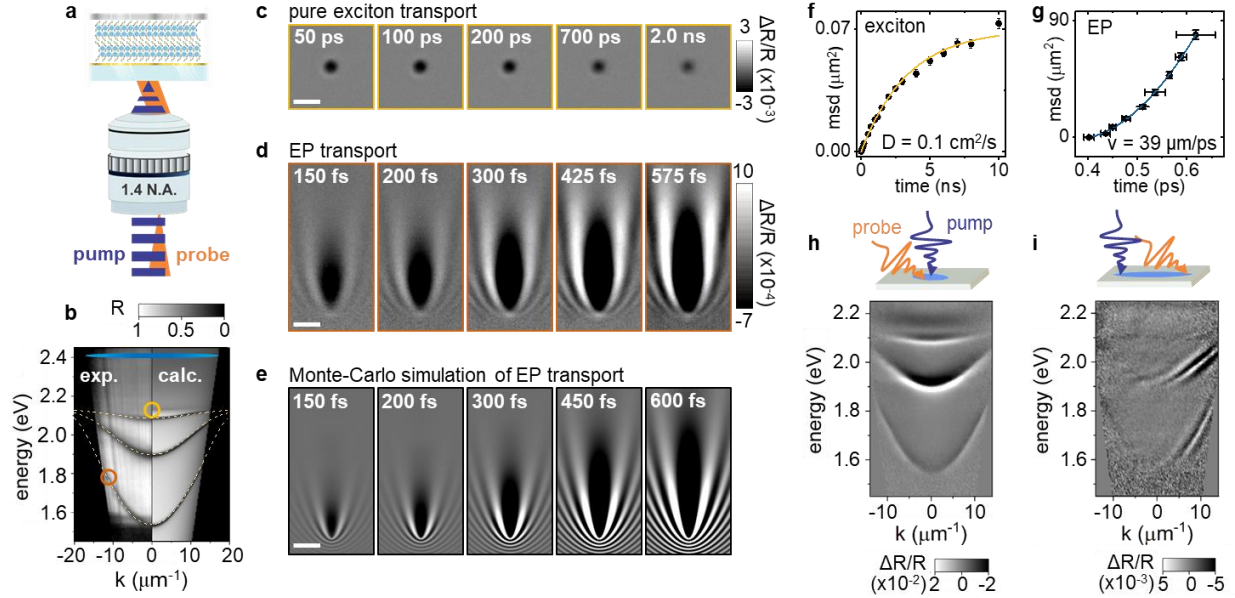

**Figure 1. Tracking EPs in a layered halide perovskite microcavity.** (a) MUPI setup and sample. (b) Momentum-resolved white light reflectance spectrum of our primary sample, a 0.67 μm thick slab of the layered halide perovskite (CH$_3$(CH$_2$)$_3$NH$_3$)$_2$(CH$_3$NH$_3$)Pb$_2$I$_7$ flanked by two metallic mirrors. Dashed lines correspond to a coupled oscillator model fit; the right side of the figure is a scattering matrix simulation of the structure (Figure S4). (c) Exciton transport probed at $k = 0$, $E$ = 2.17 eV following above-gap (2.41 eV) pump excitation. (d) EP transport probed at $k = 8.98$ μm$^{-1}$, $E = 1.77$ eV following 2.41 eV pump excitation. The probe energies and momenta used for panels (c) and (d) are illustrated with circles in panel (b); the pump energy and momentum range is illustrated with a blue ellipse. (e) Monte-Carlo simulation of MUPI contrast generated during EP propagation. Scale bars for panels (c-e) are 2 μm. (f) Mean squared displacement (msd) of bare excitons from data in panel (c). Error bars are one standard deviation. The solid curve is a fit assuming trap-limited diffusive transport. (g) msd of EPs from data in panel (d). Error bars are one standard deviation. The solid curve is a fit assuming ballistic transport. (h) Differential pump ON/pump OFF angle-resolved reflectance spectrum obtained at 1 picosecond pump-probe time delay, displaying pump-induced modification to the EP dispersion when pump and probe beams are spatially overlapped. (i) Same as panel (h), with the probe spatially separated from the pump by 1.1 μm, selectively probing EP species that have propagated away from the excitation spot.

We focus our analysis on a 0.67 μm thick slab of the layered halide perovskite (CH$_3$(CH$_2$)$_3$NH$_3$)$_2$(CH$_3$NH$_3$)Pb$_2$I$_7$ flanked by two metallic mirrors (Figure S3a). The dispersion of our structure is shown in Figure 1b, with three resolved lower polariton (LP) branches. A fit to the experimental dispersion using a coupled oscillator model indicates a Rabi splitting of 275 meV (Figure S4), similar to previous reports[32]. Figure 1c displays excitonic transport in this structure by pumping above-gap and probing at resonance with the exciton reservoir at $k = 0$ (see also Supplementary Movie 1 and analysis in Figure S6; the probe energy and momentum are highlighted with the orange circle in Figure 1b). Figure 1f plots the exciton mean squared displacement (msd), defined as msd = $\sigma^2(t) - \sigma^2(0)$, where $\sigma$ is the Gaussian width of the population profile, and $t$ is the pump-probe time delay (Supplementary Note 1). The exciton msd is exponentially-decaying over a few nanoseconds, characteristic of trap-limited diffusive transport[34]. We extract a diffusivity of 0.10 ± 0.03 cm$^2$/s and a trap density of 29 μm$^{-2}$. These values are in good agreement with recent reports of cavity-free exciton transport in these materials[33,34], suggesting that the transport of bare excitons is unaffected by the cavity.

Figure 1d displays EP propagation at the same location using the same above-gap pump excitation conditions, but probing at an energy and momentum corresponding to the yellow circle in Figure 1b (see also Supplementary Movie 2). The MUPI image series in Figure 1d displays a fast-propagating EP signal that extends over several microns within a few hundred femtoseconds, in stark contrast to the practically static bare exciton signal of Figure 1c. The intensity of the EP signal only becomes substantial ~300 fs after non-resonant pump excitation, since the latter primarily populates excitons uncoupled to the cavity; these reservoir excitons scatter into the LP branch on timescales dictated by exciton–phonon scattering[37,38], leading to a delayed rise of the EP signal. The EP msd in Figure 1g is extracted by spatially tracking the EP wavefront (analysis in Figure S7). The msd is quadratic in time (msd $\propto t^2$, or distance $\propto t$), a characteristic signature of purely ballistic transport. The EP velocity at this momentum is 39 ± 1 μm/ps, 13% of the speed of light. The distinctive interference-like features in the MUPI image profiles arise from the tilted probe plane wave interacting with the polariton population, which we successfully reproduce in Monte-Carlo simulations (Figure 1e and Supplementary Note 2).

To elucidate optical contrast generation in MUPI, we turn to transient angle-resolved spectroscopy (Figure 1h,i). The excited state EP dispersion can be monitored any distance away from the excitation to probe how propagating *versus* non-propagating populations affect the dispersion. At the excitation spot, pump-generated excitons lead to a shift of the optical dielectric response of the perovskite semiconductor[39], which results in a uniform blueshift of all LP branches (Figure 1h, modeled in Supplementary Note 3). This exciton-induced dispersion modulation reinforces that changes to LP spectra do not necessarily reflect EP dynamics[40]. In contrast, the photoinduced dispersion 1.1 μm away from the excitation and at 1 ps pump-probe time delay (Figure 1i) reflects only the EP population, since excitons are effectively immobile on this timescale. The primary feature in Figure 1i is that LP branches are broadened compared to ground state branches (Supplementary Note 3). We attribute this broadening to self-energy renormalization as a result of EP–EP interactions[41]. In our nonlinear experiment, these interactions manifest themselves as a blockade-like effect[42], wherein the presence of pump-generated EPs precludes the probe from exciting the LP branch at exactly the same energies and momenta. "We attribute this broadening to self-energy renormalization as a result of EP–EP interactions[41]. In our nonlinear experiment, these interactions manifest themselves as a blockade-like effect[42], wherein the presence of pump-generated EPs precludes the probe from exciting the LP branch at exactly the same energies and momenta. This nonlinear interaction is momentum- and energy-specific, allowing us to directly correlate MUPI measurements of polariton transport to specific points in the dispersion, as explored below. These EP-EP interactions also generate large contrast, which we leverage to reach exceptional signal-to-noise ratios with few-minute measurements per time delay and pump fluences below 10 μJ/cm$^2$.

**Lattice phonons strongly renormalize EP propagation**. We now turn to a detailed analysis of how EP transport is affected by interactions with the material lattice. We leverage the momentum-selectivity of MUPI to directly image the transport of EPs at different parts of the dispersion, i.e. as a function of excitonic character (Figures S15-S16). By directly tracking EP propagation in the time-domain, we extract an empirical EP velocity. We then compare this velocity to the expected EP group velocity, which we extract from the gradient of the experimental dispersion ($\partial\omega/\partial k$, where $\omega$ is the angular frequency). Our key result is displayed in Figure 2a, where the data points (open symbols) are the measured velocity, and the solid lines show the expected group velocity for two EP branches. The bottom panel of Figure 2a shows results of quantum dynamical simulations that we will return to below. When the probe detuning from the exciton energy (|E-$E_{ex}$|) is large, EPs are photon-like, and the experimentally-observed propagation velocity matches the expected group velocity. As the detuning decreases, moving toward more exciton-like species, we find an increasingly large deviation from the expected group velocity.

The observed deviation of the velocity can be rationalized by exciton-mediated EP–matter scattering, an interaction that becomes stronger as the excitonic character of the EP increases[43,44]. We rule out EP–exciton and density-dependent interactions: excitons and other laser-generated species are only present in the sub-micron region defined by the laser excitation spot; yet EPs maintain constant velocities throughout the fitted propagation range, including long after they escape the photoexcited region by several microns (Figure S7). Thus, EP–matter interactions occur homogeneously in space, suggesting that EP–phonon or EP–defect scattering are responsible for the velocity renormalization. Since exciton–phonon scattering dominates the transport properties of excitons in 2D halide perovskites at room temperature[34,45,46], we infer that exciton-mediated EP–phonon scattering is the dominant contribution to the velocity renormalization. We experimentally verify this hypothesis by performing MUPI experiments at 5 K, which reduces dynamic disorder (phonon scattering) by several orders of magnitude compared to room temperature[47] while negligibly affecting static disorder. EPs at 5 K propagate at the group velocity corresponding to the experimental dispersion (Figure S17), confirming that the observed velocity renormalization at room temperature results primarily from EP–phonon scattering.

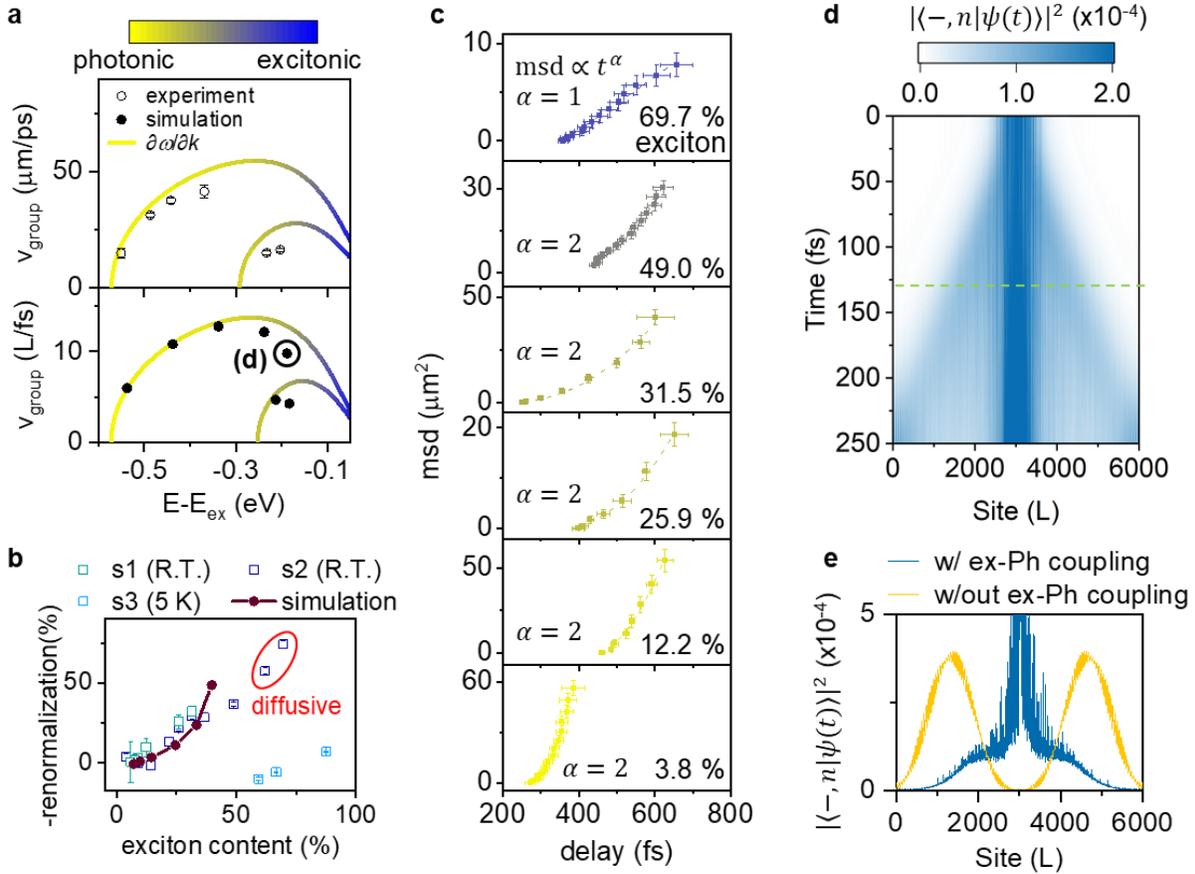

**Figure 2. EP propagation and scattering.** (a) Expected group velocity from the gradient of the dispersion (solid curve) *vs.* measured transport velocity (symbols) for each probing condition, showing an increasing deviation as exciton content is increased. $E_{ex}$ corresponds to the energy of the exciton resonance. The top panel shows results from MUPI experiments at room temperature, while the bottom panel shows results from quantum dynamical simulations. (b) EP velocity renormalization as a function of exciton content for both experimental and simulation results. Experiments are carried out at both room temperature (R.T.) and 5 K. Experiments suggest a transition to diffusive (incoherent) transport above 50% exciton content for R.T. experiments, while simulations suggest decoherence sets in above 40% exciton content. Circled data points are best fit with a diffusive ($\alpha = 1$) model (see panel c), for which velocity is not a well-defined quantity; to estimate a percentage renormalization for this data, we fit an effective velocity assuming a ballistic model ($\alpha = 2$). (c) Fits (dashed curves) to the msd extracted from MUPI experiments at R.T. Error bars are one standard deviation. (d) Quantum dynamical simulations of the spreading of the EP probability density $|\langle -, n|\psi(t)\rangle|^2$ as a function of site location (L), starting with a localized initial wave-packet with an energy-window centered at $E−E_{ex} = −0.19 ± 0.025$ eV, corresponding to the circled symbol in panel (a). (e) EP probability density at the horizontal dashed line cut of panel (d), for simulations with and without exciton-phonon coupling.

In Figure 2b, we plot the EP velocity renormalization (the percentage decrease in measured velocity compared to the expected velocity) as a function of excitonic fraction for both room temperature and cryogenic measurements. The data reveals substantial velocity renormalization of up to 40% for half-exciton EPs at room temperature. In contrast, cryogenic measurements display no renormalization for EPs with up to 88% exciton content. Data for two different room-temperature samples with different zero-momentum exciton–cavity detuning are plotted (Figures S15-S16). For a given exciton fraction, the two samples display nearly identical percentage velocity renormalizations, indicating that the latter is independent of cavity detuning. These results suggest that the velocity renormalization is primarily sensitive to the excitonic fraction, not the EP effective mass or absolute velocity. This observation is consistent with our hypothesis of exciton-mediated EP–lattice scattering. Furthermore, unlike the dispersion renormalization observed at high EP densities in the condensate regime[48], the EP dispersion itself is not renormalized in our structures. This aspect confirms that the velocity reduction we observe is not caused by density-dependent repulsive interactions or particle hybridization, but rather by scattering. Our results suggest that the commonly-used assumption that the gradient of the dispersion ($\partial\omega/\partial k$) corresponds to the polariton velocity may not be accurate for polaritons with high matter character propagating in disordered environments.

To further understand the nature of EP propagation in the presence of exciton-mediated interactions, we fit the MUPI-extracted msd to a power law, msd $\propto t^\alpha$, for EPs with different exciton fraction (Figure 2c). For exciton fractions below 50%, we observe ballistic transport ($\alpha = 2$) despite the strongly renormalized velocities. This result indicates that coherent propagation is preserved in our systems for EPs with up to 50% exciton character, since ballistic (wavelike) transport implies long-range coherence[49]. Steady-state double-slit interference measurements confirm the spatial coherence of EPs in this system (Figure S18), and the correspondence between ballistic and coherent transport is further supported by computing the purity of the density matrix of the polariton system (Figure S13). Nevertheless, above 50% exciton content, we find that the msd for room temperature samples is best fit with $\alpha = 1$, indicating diffusive (incoherent) EP propagation. In contrast, our cryogenic measurements indicate preserved ballistic transport even for EPs with up to 88% excitonic character (Figure S17). These results show that EP–phonon interactions are not just responsible for velocity renormalization, but also for the transition from ballistic to diffusive transport. Our detailed picture of EP–phonon interactions, uniquely enabled by high-resolution time-domain imaging of EP transport, reveals the rich spectrum of propagation dynamics across both photon-like and exciton-like polaritons: from ballistic and unaffected by phonons, to ballistic but severely slowed by phonon interactions, to diffusive transport. During the lengthy review process of our manuscript[50], a closely-related study performed in organic Bloch surface wave polaritons at room temperature was submitted and published; this study also displayed a group velocity renormalization and a transition from ballistic to diffusive transport[51], suggesting the behavior we are observing is general.

We turn to theory to shed more light on the nature of EP transport in the presence of dynamic disorder. Following past work on charge transport in halide perovskites[52–54], we appeal to a dynamic disorder model where transport in the absence of cavity hybridization occurs purely diffusively via the transient localization mechanism[55,56]. Our simplified, one-dimensional model Hamiltonian describes a single exciton coupled to phonons and an optical cavity (Supplementary Note 4). Coupling to the cavity has a dramatic effect on the transport properties of EPs[57]. In particular, while the instantaneous eigenstates which govern transport of the pure exciton-phonon system exhibit significant localization, polaritonic eigenstates are largely delocalized. This qualitative change shields photon-like EPs from phonon scattering, and leads to ballistic spreading of the polariton wavepacket (Figure S14). As the excitonic character of EPs increases, the percentage of localized state character correspondingly increases (non-propagating states in Figure 2d). Interestingly, however, a substantial part of the EP population continues to propagate ballistically even for large exciton content (cone-like wavepacket in Figure 2d). Although ballistic propagation is preserved, the wavepacket velocity is substantially reduced by phonon-mediated transient localization, in close agreement with our experimental observations. Figure 2e compares the computed probability distributions at 130 fs for EPs with and without exciton-phonon coupling (setting $\gamma = \alpha = 0$ in Equation S1), confirming that phonon interactions can substantially slow EP wavepackets without destroying their coherence. Our simulations predict that a high degree of coherence is maintained for up to 40% exciton fraction (Figure S13), in rough agreement with the 50% threshold observed experimentally. Our quantum dynamical simulation results are summarized in Figures 2a and 2b (filled symbols), with trends in semi-quantitative agreement with experiments despite the use of a simplified model. This correspondence provides further support to our hypothesis that group velocity renormalization and decoherence are induced by EP–phonon interactions.

**Uncoupled excitons act as reservoir states but not sinks**. The above-gap pump excitation we have used thus far generates a large population of long-lived excitons that are not coupled to the cavity, often referred to as dark states[58,59]. To elucidate the role that these excitons play in EP assemblies, we now compare transport behavior upon non-resonant (above-gap) excitation *versus* resonant excitation of the LP branch, as illustrated in Figure 3a. Figure 3b displays MUPI data when pumping above-gap and probing the LP branch up to a time delay of 1 ns; surprisingly, EP-associated signals are present 1 ns after photoexcitation, despite the EP lifetime of ~240 fs in our system (Table S1). Such long lifetimes of the LP branch are regularly observed in time-resolved spectroscopy of EP assemblies[58,60], but recent reports cast doubt on the nature of this signal[40]. The observed species propagates over many microns in less than a picosecond, establishing unambiguously that the signal corresponds to EPs. Indeed, the propagation itself ceases after a few hundred femtoseconds (limited by the intrinsic EP lifetime), but the signal persists and remains static from ~800 fs to 1 ns. These observations lend support to the exciton reservoir hypothesis[58,60], wherein uncoupled excitons populated by the non-resonant pump can scatter into the LP branch, continuously refilling the LP population throughout the exciton reservoir lifetime (6 ns in our case, Figure S5). This hypothesis is further confirmed by resonant excitation data in Figure 3c, wherein the exciton reservoir is not populated by photoexcitation. Under resonant excitation, we observe ballistic propagation of an EP wavepacket (highlighted with the orange circle), with the signal disappearing entirely after 800 fs. These results also suggest that population transfer from the LP branch to the exciton reservoir[61] is negligible in our sample.

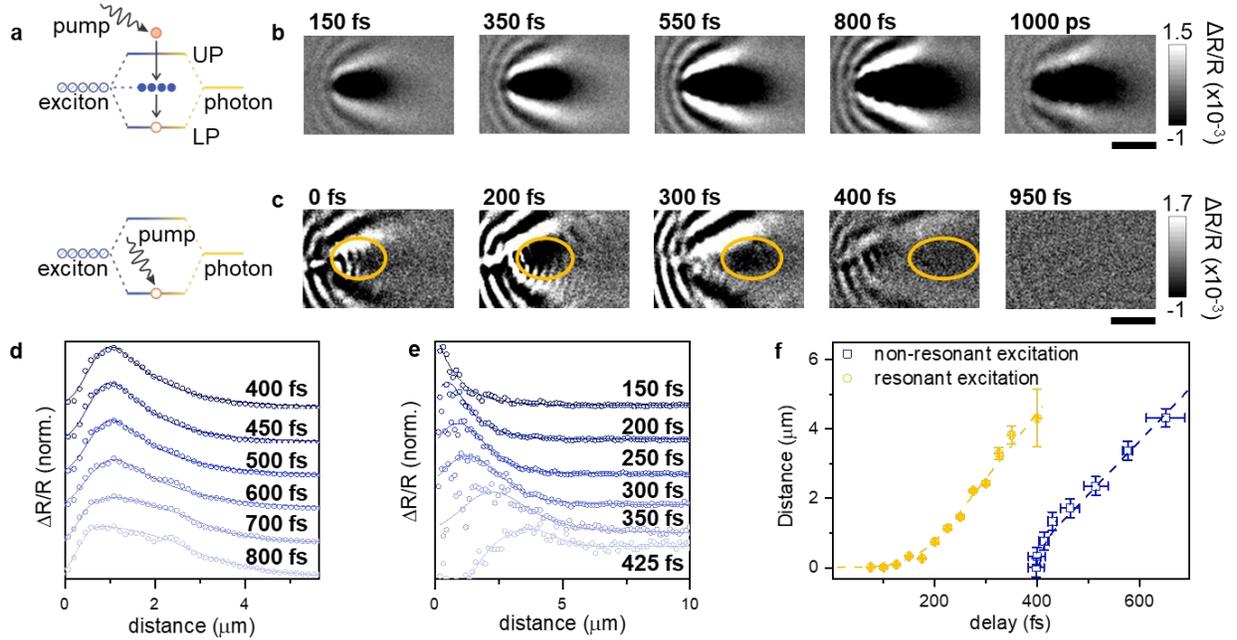

**Figure 3. Resonant vs non-resonant excitation of EPs.** (a) Tavis-Cummings model where hybridization between a cavity mode and N excitons leads to N-1 uncoupled excitons (blue circles, also known as dark states). The two different pump excitation conditions for panels (b) and (c) are illustrated at top and bottom, respectively. (b) Non-resonantly excited EPs probed at $E = 1.91$ eV and $k = 6.90$ μm$^{-1}$ (LP branch) propagate rapidly in the first ~800 fs and then remain static for a nanosecond. (c) Resonantly excited EPs with both pump and probe at $E = 1.91$ eV and $k = 6.90$ μm$^{-1}$ (LP branch) display a fast-propagating EP wavepacket that disappears after ~800 fs. The orange circle highlights the EP wavepacket on top of the strong scattering background; the latter is difficult to avoid in degenerate pump-probe microscopy. Scale bars are 2 μm. (d,e) Evolution of spatial profiles of EPs for non-resonant (d) and resonant (e) excitation. (f) EP transport extracted from the data in panels (d,e). Error bars are one standard deviation; see Supplementary Section 8 for detailed analysis. The dashed lines are from Monte-Carlo simulations of EP transport incorporating EP-EP scattering in the case of resonant excitation.

Figures 3d-f display the transport properties of non-resonantly vs resonantly-populated EPs (additional data and analysis are presented in Supplementary Note 5). In the long-time limit, both types of EPs propagate at matching velocities (Figures 3f and S19-S20), indicating that the exciton reservoir populated under non-resonant excitation does not influence the extracted EP velocities. The early-time behavior, however, is different. For the data displayed in Figure 3, non-resonant excitation leads to slow EP population buildup reaching a maximum at ~400 fs after photoexcitation; in contrast, resonant excitation leads to full population buildup within 100 fs. In the latter case, we observe slow but accelerating initial propagation; we attribute this result to the high density of EPs at the excitation location which results in strong EP–EP scattering, slowing down the initial EP propagation. As EPs decay and propagate, the EP density decreases and EP propagation speed concomitantly increases, reaching its final transport velocity ~100 fs after the signal appears. We reproduce this acceleration in Monte-Carlo simulations which take into account density-dependent EP–EP scattering (dashed lines in Figure 3f, see Supplementary Note 2 for details). These temporally inhomogeneous dynamics, similar to those recently discovered in phonon-polaritons[29], further emphasize the importance of tracking EPs throughout their lifetimes to characterize interactions between the many different component excitations of polaritonic systems.

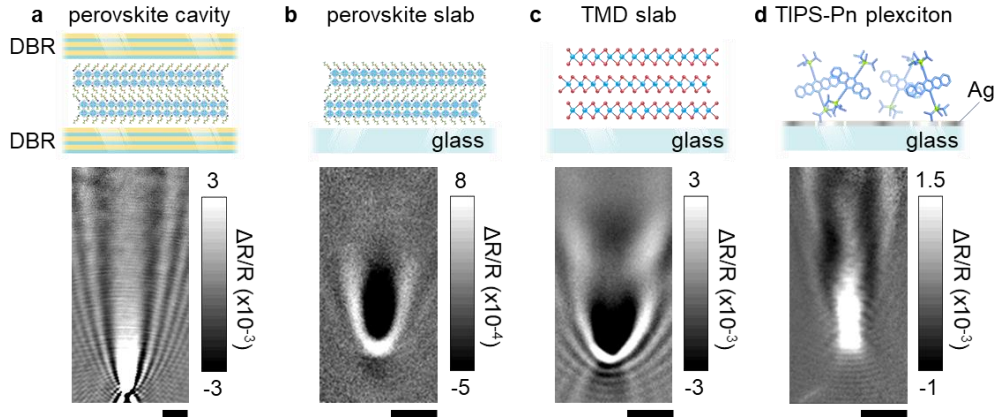

**Figure 4. Generalizability of MUPI.** Snapshots of MUPI at 1 ps pump-probe time delay following non-resonant excitation for: (a) EP transport in $CH_3(CH_2)_3NH_3)_2(CH_3NH_3)Pb_2I_7$ flanked by two Distributed Bragg Reflector (DBR) mirrors; (b) EP transport in a 1.13 μm thick layered halide perovskite slab $(CH_3(CH_2)_3NH_3)_2(CH_3NH_3)Pb_2I_7$ with no artificial cavity (Figure S21a); (c) EP transport in a 69 nm thick flake of $WSe_2$ with no artificial cavity (Figure S21b); (d) Plexciton transport in a 30 nm Ag film/50 nm TIPS-Pn amorphous film heterostructure. Scale bars are all 2 μm.

**Generalizability of MUPI.** Finally, we show that MUPI is generalizable to many different polaritonic assemblies, though we do not perform a detailed analysis here. Figure 4 displays MUPI snapshots of propagating polaritons at 1 picosecond delay for four different systems: (a) Cavity polaritons in a Distributed Bragg Reflector (DBR) cavity of a layered perovskite, showing long-range transport beyond the 19 μm field of view; (b) Self-hybridized EPs (no artificial cavity) in the same perovskite[32]; (c) Self-hybridized EPs in a transition metal dichalcogenide ($WSe_2$) slab[27,62]; and (d) Plasmon-exciton polaritons (plexcitons) in systems comprised of amorphous 6,13-Bis(triisopropylsilylethynyl)pentacene (TIPS-Pn) deposited on plasmonic silver thin films. We observe ballistic transport on femtosecond scales (except for plexcitons where the full propagation occurs within our instrument response), followed by static signals over more than tens of picoseconds when non-resonant excitation is used. As expected, propagation distances in self-hybridized cavities are much shorter compared to artificial cavities due to the shorter EP lifetime. Plexcitons exhibit much longer propagation lengths thanks to the highly dispersive plasmon modes imparting group velocities approaching the speed of light, despite the disordered nature of the amorphous molecular system used here.

In conclusion, we have developed a momentum-selective ultrafast optical imaging approach that directly visualizes EP propagation in real space and time in a wide range of emerging semiconductors. Importantly, we find that the group velocities of EPs with large excitonic character are substantially renormalized through scattering with the material lattice, which we attribute primarily to EP–phonon scattering in our layered halide perovskite microcavities. These results indicate that the commonly-used assumption that the polariton velocity corresponds to the gradient of the dispersion may not be a good approximation, particularly for highly excitonic polaritons. Remarkably, however, EPs maintain ballistic transport even in these strongly-interacting environments for up to half-exciton EPs. For EPs with excitonic fractions higher than 50%, we observe diffusive, incoherent transport at room temperature. Quantum dynamical simulations and cryogenic measurements indicate that the transition from non-interacting, to ballistic with renormalized velocities, to incoherent transport arises from the interplay of transient localization of EPs induced by strong exciton–phonon interactions and partial shielding of these interactions by hybridization with the cavity. Overall, we have established a general framework that enables precisely balancing EP coherence, velocity, and nonlinear interactions for any given polaritonic architecture. We believe these measurements will be crucial to optimize next-generation polaritonic technologies that seek to truly harness the best of their light and matter components, for example to incorporate single-photon gates[11,12,14] into scalable quantum circuits requiring long-range propagation.

**Methods**

Details of sample synthesis, cavity fabrication, optical measurements and theory are provided in the Supplementary Information.

A schematic of MUPI is shown in Figure S1. A 40 W Yb:KGW ultrafast regenerative amplifier (Light Conversion Carbide, 40W, 1030 nm fundamental, 1 MHz repetition rate) seeds an optical parametric amplifier (OPA, Light Conversion, Orpheus-F) with a signal tuning range of 640 – 940 nm and an average pulsewidth of 60 fs. For non-resonant excitation experiments, the second harmonic of the fundamental (515 nm) is used as a pump pulse, and the OPA signal is used as probe. For resonant excitation (single-color) experiments, the OPA signal is split in pump and probe beams. Group delay dispersion is partially pre-compensated using a pair of chirped mirrors (Venteon DCM7). The pump pulse is sent collimated into a high numerical-aperture objective (Leica HC Plan Apo 63x, 1.4 NA oil immersion), resulting in diffraction-limited excitation on the sample. Typical pump fluence incident on the semiconductor is 5 μJ/cm$^2$. The probe is sent to a computer-controlled mechanical delay line for control over pump-probe time delay, and is combined with the pump beam through a dichroic mirror. An f = 250 mm widefield lens is inserted prior to the dichroic mirror to focus the probe in the back focal plane of the objective for widefield illumination of the sample. A tilting mirror placed one focal length upstream of the widefield lens allows tuning the angle at which the widefield probe illuminates the sample, thus allowing probing at any momentum up to a maximum of k/k$_0$ = 1.4, limited by the numerical aperture of the objective.

Backscattered light from the sample is collected through the objective, directing the light to two different detection paths. For angle-resolved linear and transient reflectance (Figure 1 h-j), the back focal plane of the objective is projected on the entrance slit of a home-built prism spectrometer using a pair of lenses ($f_1$ = 300 mm and $f_2$ = 100 mm). For real-space MUPI imaging, this projected back focal plane image is Fourier transformed again into real-space using a 150 mm lens, forming an image on a CMOS camera (Blackfly S USB3, BFS-U3-28S5M-C). Both the spectrometer camera and the real-space camera are triggered at double the pump modulation rate, allowing the consecutive acquisition of images with the pump ON followed by the pump OFF.

**Data availability**
All raw data are displayed in Figures 1-4 of the main text and Figures S2-S21 of the Supplementary Information. Raw image files are available from the corresponding authors upon request.

**Code availability**
The source code for quantum dynamics simulations is available at[63]
https://github.com/arkajitmandal/BalisticPolaritons

**Acknowledgements**

This material is based upon work supported by the National Science Foundation under Grant Numbers DMR-2115625 (M.D.) and CHE-1954791 (D.R.R.). Revisions were primarily supported by the National Science Foundation under Grant Number CHE-2203844 (M.D.). M.D. also acknowledges support from the Arnold and Mabel Beckman Foundation through a Beckman Young Investigator award. Synthesis of WSe$_2$ (S.L.) was supported by the NSF MRSEC program through Columbia in the Center for Precision-Assembled Quantum Materials (DMR-2011738). This work used the Extreme Science and Engineering Discovery Environment (XSEDE), which is supported by National Science Foundation grant number ACI-1548562 (allocations: TG-CHE210085). Specifically, it used the services provided by the OSG Consortium, which is supported by the National Science Foundation awards #2030508 and #1836650.

**Author contributions**

D.X. and M.D. conceived and designed the experiments. D.X. and J.B. developed the instrument. D.X, S.C., I.L. and H.S. acquired the experimental data. S.L. synthesized WSe$_2$ crystals. D.X. analyzed the data. A.M. and D.R.R. developed and performed the quantum dynamical simulations. D.X., A.M., D.R.R. and M.D. wrote the manuscript, with input from all authors.

**Competing interests**

The authors declare no competing interests.

**Additional Information**

Correspondence and requests for materials should be addressed to David Reichman (drr2103@columbia.edu) and Milan Delor (milan.delor@columbia.edu)

Supplementary information for

# Ultrafast imaging of polariton propagation and interactions

Ding Xu[1†], Arkajit Mandal[1†], James M. Baxter[1], Shan-Wen Cheng[1], Inki Lee[1], Haowen Su[1], Song Liu[2], David R. Reichman[1*], Milan Delor[1*]

1. Department of Chemistry, Columbia University, New York, NY 10027, United States
2. Department of Mechanical Engineering, Columbia University, New York, NY 10027, United States

[†] These authors contributed equally

# Content

**Supplementary Methods**

**Sample preparation**

**CH$_3$(CH$_2$)$_3$NH$_3$)$_2$(CH$_3$NH$_3$)Pb$_2$I$_7$ (BA$_2$(MA)Pb$_2$I$_7$) crystal synthesis**. Ruddlesden-Popper perovskites (RPP)[1] have a general formula, A$_2$A'$_{n-1}$M$_n$X$_{3n+1}$, where A and A' are cations, M is a metal, X is a halide, and $n$ is the perovskite layer thickness. The preparation of EP cavities for MUPI measurements requires exfoliating large-area high-quality flakes of uniform thickness, which we could achieve for $n = 1$ and $n = 2$ flakes but was more challenging for $n > 2$ RPP perovskites. We focus the text on $n = 2$ flakes (BA$_2$(MA)Pb$_2$I$_7$). All chemicals were purchased from Sigma-Aldrich. The synthesis followed a published procedure.[2] Hydriodic acid (HI) solution was prepared by mixing 57% wt. aqueous HI (9 mL) and 50% wt. aqueous Hypophosphorous acid (H$_3$PO$_2$, 1 mL). Lead (II) iodide (PbI$_2$, 99.999% trace metals bases) powder (2720 mg, 5.9 mmol) was dissolved in the mixture solution under constant magnetic stirring. Methylammonium iodide (MAI, ≥ 99%, anhydrous 493 mg, 3.1 mmol) was added to the solution and black powder precipitated instantly. The black powder redissolved quickly by heating the solution to 100 °C with stirring. Subsequent addition of n-Butylammonium iodide (BAI, 864 mg, 4.3 mmol) caused the crystallization of bright orange flakes on the top of the solution. The solution was heated up to 105 °C under constant magnetic stirring until all precipitation dissolved. The solution was subjected to controlled cooling rate of 0.5 °C/h to room temperature in an oil bath, and large crystals formed on the solution surface. The crystals were collected by vacuum filtering and washed twice with toluene.

**Metallic reflectors cavity fabrication.** The bottom partial reflector through which light impinges the sample is a 30 nm gold film that was deposited on cover glass (Richard-Allan Scientific, 24×50 #1.5) by e-beam evaporation (Angstrom EvoVac deposition system). The deposition rate was 0.05 nm/s. The perovskites flakes were then mechanically exfoliated onto the gold-deposited cover glass with PVC tape (Nitto SPV224 PVC Vinyl Surface Protection Specialty Tape). Although gold is more lossy compared to silver, thin (<50 nm) silver films tend to deteriorate during measurements, which rapidly leads to poor imaging quality, an important aspect for our scattering-based experiments. The top reflector is a 200 nm silver film that was prepared by e-beam deposition on a silicon wafer. A layer of polyvinylpyrrolidone (PVP) solution (Sigma Aldrich, M.W. 40000, 10% wt in ethanol/acetonitrile wt 1/1) was then spin-coated on the silver film (3000 rpm, acceleration 1000rpm/s, 2min) and thermally annealed at 150 °C for 5 min. The prepared PVP/Ag was picked up with thermal release tape (semiconductor corp., release temperature 90 °C) and placed on the perovskites flakes with firm pressure, completing the cavity structure. The full structures were encapsulated between glass slides using epoxy (OG159-2, Epoxy Technology) in a nitrogen-filled glovebox to prevent sample oxidation during the measurements.

**Distributed Bragg reflectors (DBR) cavity fabrication.** 16/8 pairs of SiO$_2$/Si$_3$N$_4$ (107 nm/75 nm) were deposited on coverslips (Richard-Allan Scientific, 24×50 #1.5) to fabricate the top/bottom mirrors, respectively. The deposition was performed by plasma enhanced chemical vapor deposition (PECVD, OXFORD PlasmaPro®NPG80) under a pressure of 15 mTorr and a radio-frequency (RF) power of 60 W. The coverslips substrates were kept at 300 °C during the deposition. The perovskites solution was prepared by dissolving the 10 mg of crystalized (BA)$_2$MAPb$_2$I$_7$ in 5 mL N, N-dimethylformamide (Thermo Scientific™, anhydrous, 99.8%). Subsequently, $(BA)_2MAPb_2I_7$ slabs are grown between two DBR substrates by slow solvent evaporation in an Argon-protected atmospheres at 60 °C for an hour.

**$WSe_2$ sample preparation**. The self-hybridized $WSe_2$ cavity sample was prepared by direct Scotch-tape mechanical exfoliation of bulk high-purity flux-grown $WSe_2$ crystals[3,4] onto a borosilicate cover glass. Flakes of the right thickness were found using atomic force microscopy (Bruker Dimension FastScan Atomic Force Microscope).

**6,13-Bis(triisopropylsilylethynyl)pentacene (TIPS-pentacene) plexciton sample preparation.** A 30 nm silver film was deposited on cover glass by e-beam deposition (at 0.05 nm/s), creating a plasmonic film. The silver is then coated with a 3 nm $Al_2O_3$ top layer that acts to both protect the silver and prevent direct charge-transfer interactions with molecules. $Al_2O_3$ that was grown by atomic layer deposition (SAVANNAH 200, Cambridge Nano Tech Inc.). TIPS-Pentacene (≥ 99%, 20 mg) and poly(methyl methacrylate) (PMMA, M.W. 35000, 6 mg) was dissolved in toluene (1 mL) under constant stirring. The TIPS-pentacene/PMMA solution is then spin-coated on the silver-coated cover glass at 3000 rpm for 60 seconds.

**Optical measurements: Momentum-resolved Ultrafast Polariton Imaging**

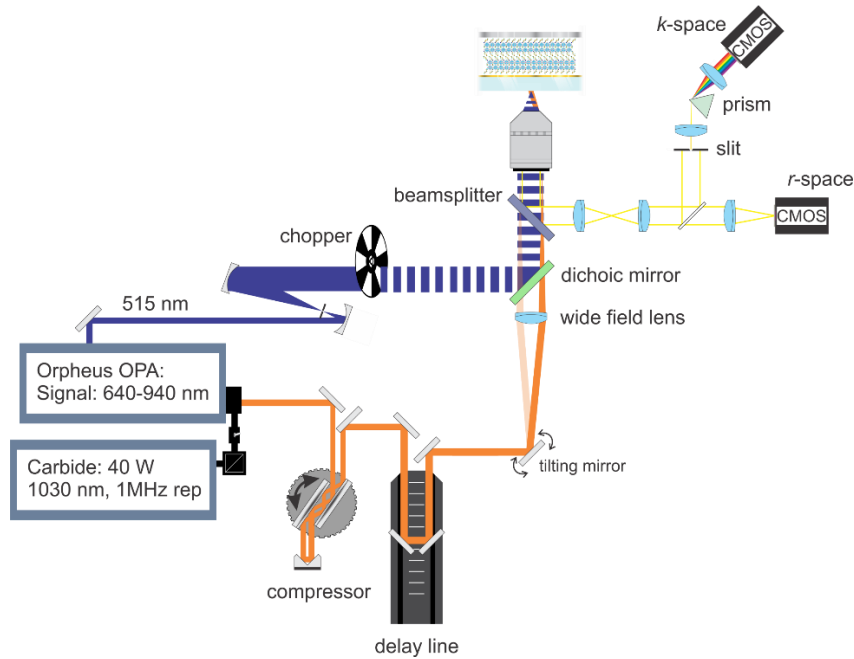

**Figure S1. MUPI schematic.** Optical schematic for momentum-resolved ultrafast polariton imaging.

MUPI is shown in Figure S1. A 40 W Yb:KGW ultrafast regenerative amplifier (Light Conversion Carbide, 40W, 1030 nm fundamental, 1 MHz repetition rate) seeds an optical parametric amplifier (OPA, Light Conversion, Orpheus-F) with a signal tuning range of 640 – 940 nm and an average pulsewidth of 60 fs. For non-resonant excitation experiments, the second harmonic of the fundamental (515 nm) is used as a pump pulse, and the OPA signal is used as probe. For resonant excitation (single-color) experiments, the OPA signal is split in pump and probe beams using a beamsplitter. Dispersion of the OPA signal caused by refractive optics (in particular the microscope objective) is partially pre-compensated using a pair of chirped mirrors (Venteon DCM7). For all experimental configurations, the pump pulse train is modulated at 647 Hz using an optical chopper and is sent collimated into a high numerical-aperture objective (Leica HC Plan Apo 63x, 1.4 NA oil immersion), resulting in diffraction-limited excitation on the sample. Typical pump fluence incident on the semiconductor is 5 μJ/cm$^2$. The probe is sent to a computer-controlled mechanical delay line for control over pump-probe time delay, and is combined with the pump beam through a dichroic mirror. An f = 250 mm widefield lens is inserted prior to the dichroic mirror to focus the probe in the back focal plane of the objective, resulting in widefield illumination of the sample. A tilting mirror placed one focal length upstream of the widefield lens allows tuning the angle at which the widefield probe illuminates the sample, thus allowing probing at any momentum up to a maximum of $k/k_0 = 1.4$, limited by the numerical aperture of the objective.

A beamsplitter collects the backscattered light from the sample through the objective, directing the light to two different detection paths. For angle-resolved linear and transient reflectance (Figure 1 h-j of the main text), the back focal plane of the objective is projected on the entrance slit of a home-built prism spectrometer using a pair of lenses ($f_1$ = 300 mm and $f_2$ = 100 mm), as depicted in Figure S1. For real-space MUPI imaging, this projected back focal plane image is Fourier transformed again into real-space using a 150 mm lens, forming an image on a CMOS camera (Blackfly S USB3, BFS-U3-28S5M-C). Both the spectrometer camera and the real-space camera are triggered at double the pump modulation rate, allowing the consecutive acquisition of images with the pump ON followed by the pump OFF. Consecutive frames are then processed according to (pump on/pump off – 1).

For cryogenic measurements, the same light source is used, but the sample is placed in a close-loop Montana Instruments s100 cryostation equipped with a cryo-optic objective (in-vacuum but room-temperature objective). The objective is a Zeiss LD EC Epiplan-Neofluar 100x/0.90 DIC M27 (NA = 0.9).

The configuration used for ultrafast angle-resolved transient reflectance experiments (Figure 1h,i) is almost identical to that described above, except that the probe is a supercontinuum white light generated by focusing the fundamental 1030 nm beam from the regenerative amplifier in a 5.0 mm yttrium aluminum garnet flat window (YAG, undoped, orientation[111], EKSMA Optics). For this experiment, the probe is sent collimated into the objective and overfills the back aperture of the objective. This configuration ensures that all angles allowed by the numerical aperture of the objective are collected, allowing the data in Figures 1h,i to be collected in a single pump/probe image pair, i.e. without having to scan across angles.

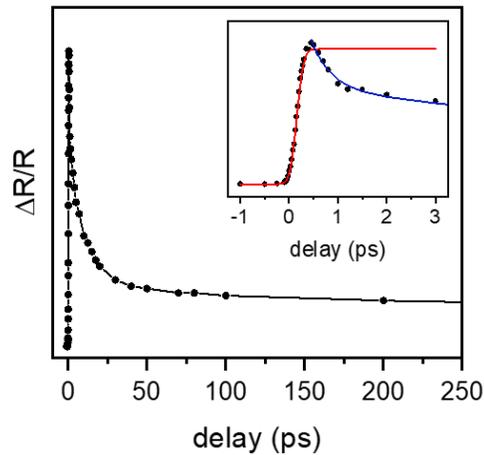

**Figure S2. Normal incident pump-probe experiment using a 515 nm pump and 730 nm probe**. The inset shows the early pump-probe time delay signal. The estimated instrument response function from the fit is 129 fs.

**Sample characterization**

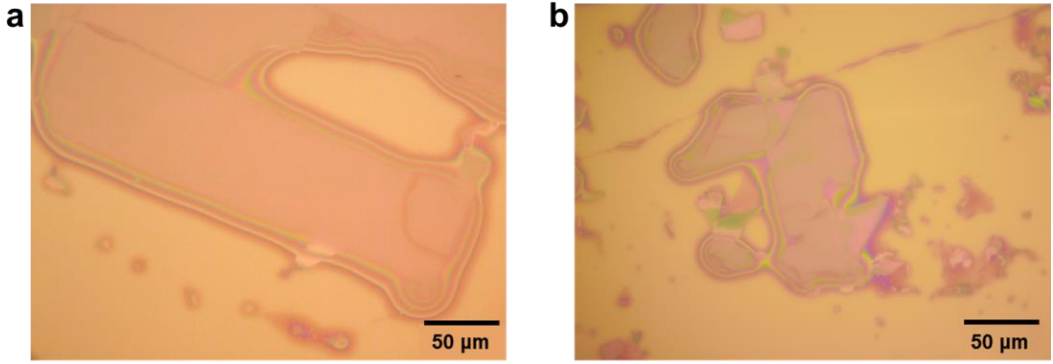

**Figure S3. Sample optical images.** Images for a 0.667 μm (a) and 0.696 μm (b) thick $BA_2(MA)Pb_2I_7$ crystal flakes enclosed in a Au-Ag cavity, corresponding to sample 1 and sample 2 in main text. This image was collected in reflection using a halogen white light lamp and a 40X objective. The scale bar is 50 μm.

**Coupled oscillator model, scattering matrix simulations, and polariton lifetimes**

As has been shown in the recent literature[5–9], multi-mode cavities comprising a semiconductor slab of finite thickness (i.e. not a monolayer) in the strong-coupling regime are best fit using a coupled oscillator model described by a 2N-dimension block-diagonal Hamiltonian for N cavity modes. For the three modes observed in the dispersion in our case, we thus model the dispersion using the following coupled oscillator model:

$$\begin{bmatrix} U_{ex} & \delta & 0 & 0 & 0 & 0 \\ \delta & H_1 & 0 & 0 & 0 & 0 \\ 0 & 0 & U_{ex} & \delta & 0 & 0 \\ 0 & 0 & \delta & H_2 & 0 & 0 \\ 0 & 0 & 0 & 0 & U_{ex} & \delta \\ 0 & 0 & 0 & 0 & \delta & H_3 \end{bmatrix} \begin{bmatrix} \chi_1 \\ \varphi_1 \\ \chi_2 \\ \varphi_2 \\ \chi_3 \\ \varphi_3 \end{bmatrix} = E_{pol} \begin{bmatrix} \chi_1 \\ \varphi_1 \\ \chi_2 \\ \varphi_2 \\ \chi_3 \\ \varphi_3 \end{bmatrix}$$

Where $U_{ex}$ is the exciton energy (2.14 eV), $H_n$ is the energy for each cavity mode, $\delta$ is the interaction energy between each cavity mode and the exciton, and $E_{pol}$ is the polariton energy. The implication of this model is that for this sample geometry, the cavity modes don't interact with each other. As a result, each polariton branch can be modeled as a 2x2 coupled oscillator between each cavity mode and the exciton. Figure S4a (dashed lines) displays the result of such a coupled oscillator fit overlaid on the experimental dispersion for the three lower polariton branches. The fit provides a value of $\delta$ = 137.5 meV, i.e., a Rabi splitting of 275 meV. The dispersion shows S-like bending (flattening) of the lower polariton branches toward higher momenta, avoiding crossing of the exciton line – a characteristic signature of strong light-matter interaction.

Figure S4b shows the result of a scattering matrix method (SMM) calculation, performed using the open-source S4 package,[10] using the perovskite slab thickness and the known dielectric functions of the metallic mirrors and BA$_2$(MA)Pb$_2$I$_7$. The experimental dispersion, coupled oscillator model and SMM calculations are all in excellent agreement with one another. Finally, Figure S4c shows the full decomposition of the coupled oscillator model for each mode. Note that upper polariton branches are not visible in the experimental dispersion because the material is highly absorptive above bandgap, as commonly observed in the perovskite polariton literature[11,12], and as also captured in the SMM calculations (Fig. S4b).

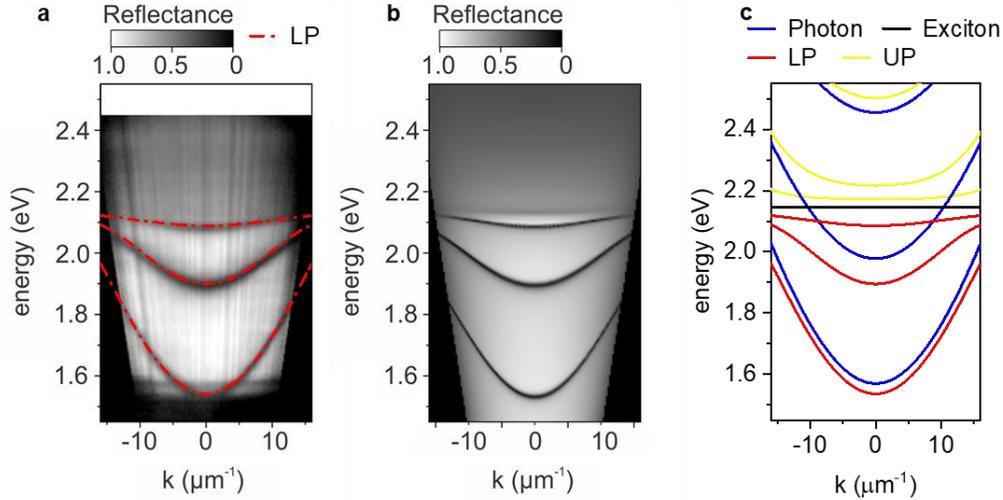

**Figure S4. Coupled oscillator and scattering matrix modeling of polariton dispersion.** (a) White light angle-resolved reflectance of the structure described in the main text (0.667 μm thick BA$_2$(MA)Pb$_2$I$_7$ flanked by two metallic mirrors). Dashed lines correspond to the lower polariton branches from a coupled oscillator model fit to the experimental dispersion. (b) Scattering matrix simulations using the empirically determined perovskite thickness and the known dielectric functions of the metallic mirrors and of BA$_2$(MA)Pb$_2$I$_7$. The simulation agrees quantitatively with the experimental dispersion. (c) Full result of the coupled oscillator model, showing the exciton, uncoupled cavity modes, and the lower and upper polariton branches.

The exciton content of the polariton for each mode is obtained by the eigenstate normalization:

$$\phi_{LP} = A\varphi - X\chi$$

$$\phi_{UP} = A\varphi + X\chi$$

$$|A|^2 + |X|^2 = 1$$

Where $A$ and $X$ are the Hopfield coefficients associated with the relative photon and exciton content of the polariton.

The EP lifetimes for the various polaritons probed in Figure 2 of the main text are estimated using two methods and tabulated in Table S1. We note that EP lifetimes are difficult to calculate exactly due to various losses and disorder not accounted for in simple models; however, the EP lifetime is not used as a parameter in any of our analyses. First, we can estimate the EP lifetime using:

$$\tau^{-1} = \frac{|X|^2}{\tau_{ex}} + \frac{1-|X|^2}{\tau_{loss}}$$

$$\tau_{loss} = \frac{2nL/\cos\theta}{1 - R_{top}(\theta)R_{bottom}(\theta)}$$

Where $\tau_{ex}$ is the exciton lifetime (~6 ns in our case, figure S5), $\tau_{loss}$ is the cavity lifetime, $L$ is the cavity thickness, $\theta$ is the angle relative to the normal of the interface, $n$ is the refractive index of the semiconductor, and $R(\theta)$ are the Fresnel reflection coefficients for the top and bottom interfaces of the cavity.

The second method we use is to calculate the lifetime through the inverse Lorentzian linewidth ($\Gamma$) of the experimental dispersion,

$$\tau = \hbar/\Gamma$$

which provides a lower bound for the polariton lifetime. Most lifetimes range from ~100 – 300 fs. For completeness, we also calculate the inverse linewidth obtained from SMM calculations, which provides an upper lifetime limit assuming no disorder and losses not accounted for in the dielectric functions of the system.

Table S1: Estimated polariton lifetimes for different exciton content for the two samples (at room temperature) described in Figure 2.

| | polariton energy (eV) | exciton content (%) | Calculated cavity photon lifetime $\tau_{loss}$ (fs) | Calculated polariton lifetime $\tau$ (fs) | Inverse SMM linewidth lifetime (fs) | Inverse experimental linewidth lifetime (fs) |
|---|---|---|---|---|---|---|
| 1 | 1.937 | 31.5 | 174 | 256 | 306 | 149 |
| | 1.907 | 25.9 | 166 | 225 | 263 | 159 |
| | 1.771 | 12.2 | 199 | 227 | 489 | 138 |
| | 1.698 | 8.8 | 177 | 195 | 318 | 132 |
| | 1.653 | 7.4 | 165 | 179 | 400 | 87 |
| | 1.590 | 5.9 | 151 | 161 | 295 | 135 |
| 2 | 2.049 | 69.7 | 328 | 1083 | 815 | 299 |
| | 2.033 | 62.1 | 288 | 758 | 689 | 260 |
| | 2.000 | 49.0 | 256 | 481 | 430 | 193 |
| | 1.959 | 36.5 | 204 | 321 | 403 | 187 |
| | 1.937 | 31.5 | 181 | 264 | 307 | 154 |
| | 1.907 | 25.9 | 178 | 240 | 264 | 146 |
| | 1.879 | 21.7 | 130 | 167 | 246 | 140 |
| | 1.805 | 14.4 | 197 | 231 | 475 | 95 |
| | 1.698 | 8.8 | 155 | 170 | 318 | 75 |
| | 1.442 | 3.8 | 181 | 188 | 255 | 72 |

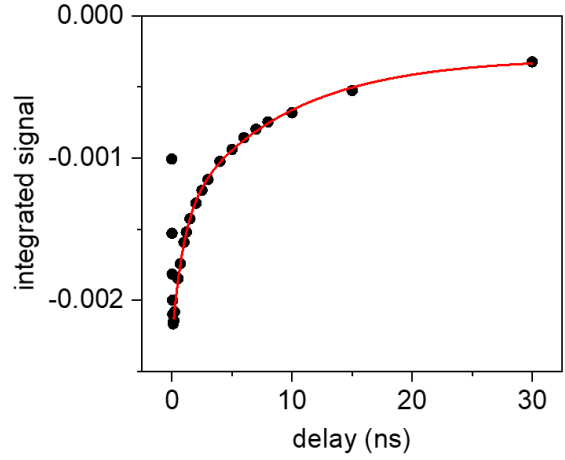

**Figure S5. Time-resolved pump-probe experiment with pump at 440 nm and probe at 570 nm to probe the bare exciton lifetime**. The red line is a biexponential fit with lifetimes of 0.65 and 6.11 ns.

Finally, the expected group velocity $v_g$ of the polariton is calculated by first-order differentiation of the dispersion relation:

$$v_g = \frac{\partial \omega}{\partial k}$$

Where $\omega$ is the resonance (angular) frequency of the polariton and $k$ is the momentum.

**Supplementary Note 1**

**Exciton and EP transport analysis and spatial precision**

The mean squared displacement (msd) of the photoexcited species in spatiotemporal microscopies is defined as:[13–15]

$$\text{msd} = \sigma^2(t) - \sigma^2(0) = 2Dt^\alpha$$

where $\sigma$ is the Gaussian width, $t$ is the pump-probe time delay, $D$ is the diffusivity, and $\alpha$ is an exponent characterizing the transport regime. For diffusive transport, $\alpha = 1$; for sub-diffusive transport, $\alpha < 1$; in the limit of ballistic transport, $\alpha = 2$ (corresponding to distance $\propto$ time).[14,15] Thus, the msd enables us to unambiguously characterize different transport regimes observed for excitons and polaritons. The spatial precision with which exciton or polariton transport can be characterized is not defined by the diffraction limit, but rather by the signal-to-noise ratio of the measurement[14]. For example, we can define a propagation length $L = \sqrt{\text{msd}}$ which has an error

$$\Delta L = \sqrt{\frac{\sigma^2(t)}{\sigma^2(t) - \sigma^2(0)} \Delta \sigma^2(t) + \frac{\sigma^2(0)}{\sigma^2(t) - \sigma^2(0)} \Delta \sigma^2(0)},$$

This error is the spatial precision of our measurement, and is set by the precision with which the variance can be fit. Our measurements typically yield sub-30 nm spatial precision.

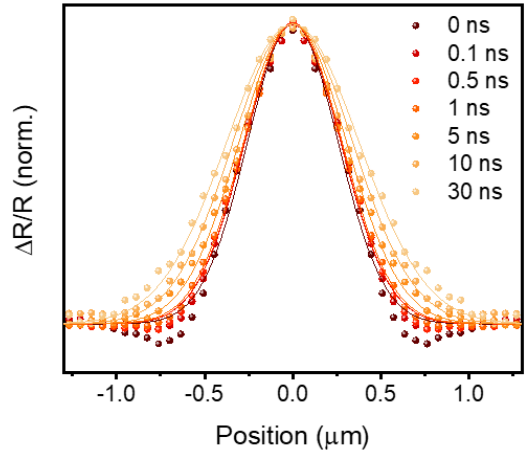

**Figure S6. Bare exciton transport in BA$_2$(MA)Pb$_2$I$_7$ taken with a probe at the exciton resonance at $k = 0$.** The population profiles were obtained through radial averaging of the data shown in Figure 1c of the main text. Gaussian fits to the population profiles allow extracting $\sigma$ and thus the msd. In our 2D halide perovskites, exciton transport is observed to be sub-diffusive, and is fit to a trap-limited, exponentially-decaying diffusivity in agreement with recent reports[16], as detailed in the main text. The pump fluence for this dataset is 4.44 μJ/cm$^2$.

For polaritons probed at finite $k$, the propagation is one-sided; the variance $\sigma^2(t)$ can either be obtained by a single-sided Gaussian fit to the polariton profile, or by fitting the arrival time of the polariton at a specific distance away from the excitation spot, as shown in Figure S7. Both approaches give almost identical results in most datasets, but we have found the method shown in Figure S7 to be more robust in measurements with low signal-to-noise ratio.

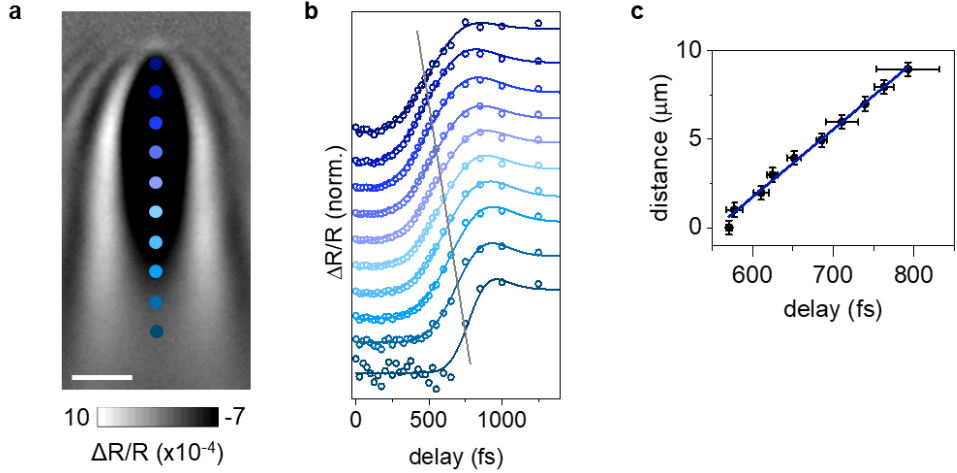

**Figure S7. Example of propagation fitting for MUPI imaging of EP transport**. Here, EPs are generated at the top of the image (a), and propagate downward. The signal rise-time for each location indicated by the blue dots in the image is extracted (panel b). Since the rise time at each location is a convolution of the instrument response function with the delayed arrival of the EP, the rise times are fit using a Gaussian function convolved with a bi-exponential decay (the second decay is a nanosecond decay component to account for the offset),

$$\frac{\Delta R}{R} = A_1 e^{-\frac{t}{\tau_1}+\frac{t_0+w^2}{2\tau_1^2}}\left(1+\text{erf}\left(\frac{t-t_0-\frac{w^2}{\tau_1}}{\sqrt{2}w}\right)\right) + A_2 e^{-\frac{t}{\tau_2}+\frac{t_0+w^2}{2\tau_2^2}}\left(1+\text{erf}\left(\frac{t-t_0-\frac{w^2}{\tau_2}}{\sqrt{2}w}\right)\right) + C$$

where $A_1, A_2, \tau_1, \tau_2$ are fit amplitudes and decay times, $w$ is the instrument response function width, and $t_0$ is defined as the EP arrival time. Additional fits and datasets are provided in Figure S15-17. Note that this fitting procedure provides almost identical results to extracting the EP arrival time as the time at which the EP signal rises to half its maximum amplitude (schematically illustrated with the black line in panel b). (c) Resulting distance vs delay plot extracted from the fitting procedure, showing linear behavior. Note that the msd (rather than distance) is plotted in the main text. Error bars are one standard deviation

**Supplementary Note 2**

**Monte-Carlo simulations of polariton transport**

To model the imaging patterns observed in MUPI, Monte-Carlo simulations such as that shown in Figure 1e of the main text were carried out. In these simulations, imaged EP 'particles' are assumed to move ballistically at the group velocity extracted from the polariton dispersion, and along the wavevector probed by the probe field, until they are elastically scattered by the lattice. For non-resonant excitation, the EP injection rate is modeled according to the empirical rise-time of the signal; this finite rise-time reflects population scattering from the long-lived exciton reservoir to the lower polariton branch (Figure S8). For resonant excitation, the EP injection rate is modeled after the instrument response function (Figure S2). In the Monte-Carlo model, the EP particles are initialized having the same group velocity, and with a spatial Gaussian distribution

$$G(r, \sigma) = \frac{1}{\sqrt{2\pi}\sigma} \exp\left(-\frac{r^2}{2\sigma^2}\right)$$

where $r$ is the distance of the EPs to the excitation origin and $\sigma$ is the standard deviation of the distribution, which is set as the diffraction limit for the pump wavelength used. The directions of EPs are randomized in the initialization. At each time step of dt = 10 fs, we calculate the probability of EP-lattice scattering and EP loss with $\exp(-dt/\tau_{phonon})$ and $\exp(-dt/\tau)$, where $\tau_{phonon}$ and $\tau$ are the EP-lattice scattering time and polariton lifetime, respectively. In these simulations, $\tau_{phonon}$ is empirically tuned to match the transport behavior observed in our experiments. The elastically scattered EPs are then assigned a new random angle (Figure S9a). To reproduce MUPI images, wherein the probe selects only EPs with specific wavevectors, we selectively plot EPs populated within a finite range of angles, even though the simulation includes all EPs with different momenta. This angle range is extracted from the $k$-width of the EP dispersion curve in the angle-resolved reflectance spectra, fitted by a Lorentzian function. All simulations are averaged over 120 independent runs.

For EP simulations of resonant excitation (Figure 3f), where the EP population can saturate within the pump pulse width, we additionally introduce EP-EP scattering. We estimate the scattering time using the semiclassical approximation[17]:

$$\tau_{scatter} = (\tau_{pol}^{-1} + \tau_{phonon}^{-1})^{-1}$$

$$\tau_{pol}^{-1} = \sigma n v$$

where $\sigma$ is the EP scattering cross-section, $n$ is the polariton density, and $v$ is the group velocity. $\sigma$ was calculated by $\sigma = \sigma_{EX}|X|^4$, where $\sigma_{EX}$ is the exciton-exciton scattering cross-section (4 nm, assigned as the exciton Bohr radius[18]).

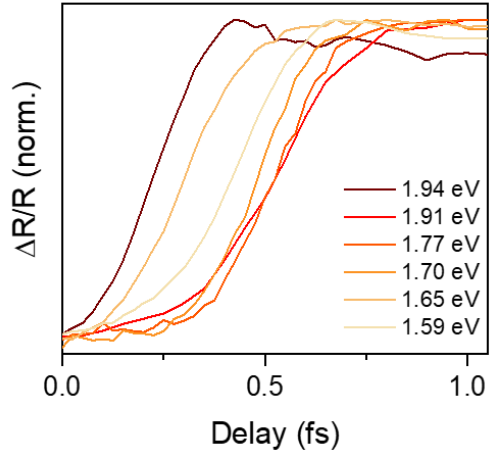

Figure S8. The temporal rise-time for EPs probed at different energies (corresponding to the data for sample 1 in Figure 2) following above-gap excitation.

To model the MUPI image formation process, the spatial EP population distribution is convolved with a filter function that models the interference between the widefield probe (plane wave) and spherical wave scattering from the EP population (Figure S9), i.e.:

$$\mathbf{E}_{sp} = \frac{A}{r} e^{i(k|\mathbf{r}-\mathbf{r}_1|-\omega t + \phi_1)}$$

$$\mathbf{E}_p = C e^{i(\mathbf{kr}-\omega t + \phi_2)}$$

$$I = |\mathbf{E}_{sp} + \mathbf{E}_p|^2 = |\mathbf{E}_{sp}|^2 + |\mathbf{E}_p|^2 + 2\mathbf{E}_{sp} \cdot \mathbf{E}_p = \left(\frac{A}{r}\right)^2 + C^2 + 2\frac{AC}{r} \cos[k(r - r\cos\theta) + \Delta\phi]$$

Where $I$ is the light intensity detected on the CMOS camera, A and C are the wave amplitudes, $k$, ω are respectively the spatial angular frequency and circular frequency of the electromagnetic wave. $\mathbf{k}$ is the wavevector, $\mathbf{r}$ is the position vector, and $\theta$ represents the angle between the vectors. $\phi$ is an arbitrary phase tuned to match the experimental profile. Subscripts p and sp correspond to plane wave and spherical wave, respectively. The resulting images (shown in Figure 1e of the main text) for different time delays closely reproduce the characteristic MUPI profile of interference-like features near the pump excitation location, as well as a ballistically expanding wavefront along the probed wavevector.

All Monte-Carlo simulations were carried out in python.

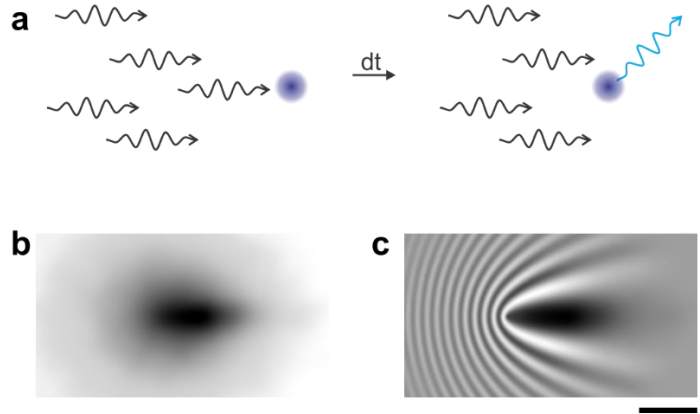

**Figure S9. Monte-Carlo simulations of EP transport**. (a) Illustration of the time-stepping process in our simulations. (b) The spatial distribution of EPs as 'particles' in the Monte-Carlo simulations at a time delay of 400 fs, for EPs at 1.91 eV with a group velocity of 20.35 µm/ps and a measured propagation speed of 15.3 µm/ps. The black scale level represents EP density. (c) Same as panel (b) after convolution with the interference filter function. The scale bar is 2 µm.

## Supplementary Note 3

**Simulations of ground and transient angle-resolved reflectance spectra**

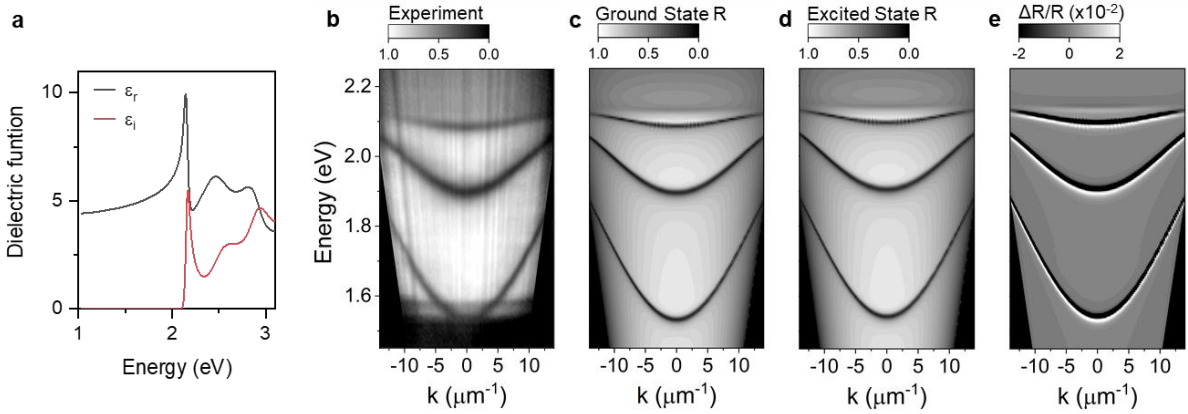

**Figure S10. Simulated linear and nonlinear angle-resolved spectra**. (a) The ground state dielectric function of $BA_2(MA)Pb_2I_7$.[19] (b) Momentum-resolved reflectance the Au/Perovskite/Ag cavity. (c) Simulated momentum-resolved reflectance spectra of the same cavity. (d) Simulated excited state reflectance spectra of the same cavity, using a blueshifted dielectric function for the 2D perovskite, and assuming 3% of oscillators are populated. (e) Transient reflectance spectra obtained by computing the differential reflectance from panel (c) and panel (d) (see text), corresponding to data presented in Figure 1h of the main text.

To simulate and analyze the angle-resolved reflectance profiles presented in Figures 1b, h, i of the main text, we turn to a scattering matrix approach, which we perform using the open-source S4 package.[10] We first start by simulating the ground state (linear) angle-resolved reflectance profile. Input simulation parameters are the ground state dielectric function of $BA_2(MA)Pb_2I_7$ (figure S10a) which we obtain from the literature;[5] incident light angle (varied from 0 to 90 degrees) and polarization (s-polarized); the cavity thickness (0.667 μm); and mirror layers (30 nm gold and 150 nm silver), with light incident through the gold. Figure S10c shows the resulting linear angle-resolved reflectance profile, which agrees well with the experimental data (Figure S10b).

The experimental transient angle-resolved reflectance profile (Figure 1h of the main text) displays a ~10 meV blueshift of all ground state polariton branches almost uniformly across all momenta. We hypothesize that this uniform blueshift arises due to a well-known pump-induced shift of the dielectric function in 2D halide perovskites. This pump-induced shift, typically on the order of 5-15 meV,[20] occurs due to excitonic many-body interactions.

To test our hypothesis, we used the scattering matrix approach to calculate the angle-resolved reflectance profile using the excited-state dielectric function of $BA_2(MA)Pb_2I_7$. Considering that not all excitons are populated, the excited state dielectric function is generated by averaging 3% of a 10 meV blueshifted dielectric with the ground state dielectric function. We then simulate the transient (differential) angle-resolved reflectance profile by calculating

$$\frac{\Delta R(\theta)}{R(\theta)} = \frac{R_{\text{excited}}(\theta) - R_{\text{ground}}(\theta)}{R_{\text{ground}}(\theta)}$$

Where $R_{ground}$ (Figure S10c) and $R_{excited}$ (Figure S10d) are generated with the ground-state and excited-state dielectric functions, respectively. As shown in Figure S10e, the simulated spectra agree closely with the experimental spectra (Figure 1h), suggesting that a simple exciton-induced dielectric-shift model is sufficient to explain the observed lower polariton blueshift for spatially overlapped pump/probe data. In other words, although the signal appears along the lower polariton branch dispersion, this signal reports on the exciton population rather than the EP population.

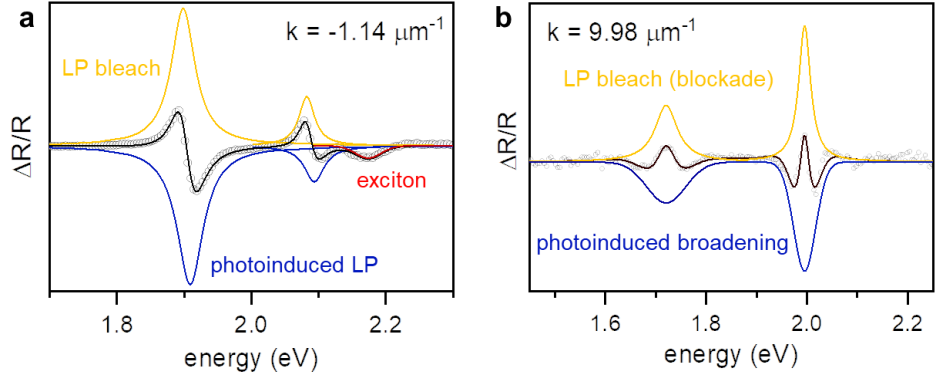

**Figure S11. Spectral cuts from transient angle-resolved reflectance spectra**. (a) spatially-overlapped pump and probe pulses, and (b): spatially-separated (>1 μm) pump and probe pulses. Data points are black circles. The black line represents the total fit, composed from adding the yellow, blue, and red (for the exciton resonance in panel a) fit components. Note that the spectral cuts are taken at different values of $k$ for the spatially-overlapped (-1.14 μm$^{-1}$) vs separated (9.98 μm$^{-1}$) pump-probe data.

To further investigate the nature of the signal in both spatially-overlapped (Figure 1h) and spatially-separated (Figure 1i) pump-probe transient reflectance profiles, we analyze spectral line cuts at specific momenta in Figure S11. For the spatially overlapped data (Figure S11a), line cuts along any momentum (here shown for $k = -1.14\ \mu m^{-1}$) provide an alternative visualization of the aforementioned dielectric blueshift simulated in Figure S10. The observed transient features are fit with two Lorentzian functions, where the positive component represents 'bleaching' of the lower polariton branch (disappearance of the lower polariton branch from its ground state energy, resulting in more reflectance), and the blueshifted negative component represents the new spectral location of the branch. The extra bump at 2.16 eV is from the exciton reservoir.

For spatially separated transient reflectance (Figure S11b), with pump probe separation distances exceeding 1 micron and at a time delay of 1 ps, the transient angle-resolved spectra must report on the EP population since excitons are immobile on this timescale. Here, the transient profiles look different: the most noticeable feature is a broadening of the polariton band. The net result is a bleach of the original polariton band (fit with a positive Lorentzian), signifying that probe photons with energies and momenta matching the ground state LP are not able to enter the cavity, and new photon pass-bands (fit with a negative Gaussian function) either side of the original polariton band, indicating that photons with slightly higher and lower energies are now able to enter the cavity and populate EPs. This behavior closely resembles a blockade-like effect[21]: when pump-populated EPs are present at a specific location in space, probe photons are unable to populate EPs at the exact same energy and location. Such repulsive EP-EP interactions are known to lead to spectral broadening (imaginary self-energy)[22], the main feature observed in Figure S11b.

These transient angle-resolved reflectance profiles thus indicate that the mechanism for contrast generation in MUPI arises from blockade-like EP-EP interactions, whereby the imaged probe field is modified by the presence of propagating pump-generated EPs: the latter change the probability for probe photons to enter the cavity at the center probe energies, resulting in enhanced or suppressed reflectivity depending on the exact probe energy and momentum selected.

## Supplementary Note 4

### Theory of EP–lattice scattering

#### 4.1 Model Hamiltonian

We consider a simple one-dimensional model[23,24] with both Peierls and Holstein-type exciton–phonon interactions coupled to a set of radiation modes inside an optical cavity[25]. The light-matter Hamiltonian in atomic units ($\hbar = 1$) is given as[25–27],

$$\hat{H} = \sum_n (\epsilon_0 + \gamma \hat{q}_n) \hat{c}_n^\dagger \hat{c}_n + \sum_n (\hat{c}_{n+1}^\dagger \hat{c}_n + \hat{c}_n^\dagger \hat{c}_{n+1})[\alpha(\hat{q}_{n+1} - \hat{q}_n) - \tau] + \sum_n \frac{\hat{p}_n^2}{2m} + \frac{1}{2} K \hat{q}_n^2$$

$$+ \sum_{\mathbf{k},n} \frac{g_c}{\sqrt{N}} \sqrt{\frac{\omega_\mathbf{k}}{\omega_0}} (\hat{\mu}_n \cdot \hat{\epsilon}_\mathbf{k}) (\hat{a}_\mathbf{k} e^{i\mathbf{k}\cdot X_n} + \hat{a}_\mathbf{k}^\dagger e^{-i\mathbf{k}\cdot X_n}) + \sum_\mathbf{k} \hat{a}_\mathbf{k}^\dagger \hat{a}_\mathbf{k} \omega_\mathbf{k} \quad (S1)$$

In the light-matter Hamiltonian expressed above, $\hat{c}_n^\dagger$ and $\hat{c}_n$ are the excitonic creation and annihilation operators at site $n = 1, 2 \ldots N$ located at $X_n = n \cdot L$ with transition dipole $\hat{\mu}_n = \hat{x}(\hat{c}_n + \hat{c}_n^\dagger)$, $\hat{a}_\mathbf{k}^\dagger$ and $\hat{a}_\mathbf{k}$ are the photonic creation and annihilation operators of cavity mode $\mathbf{k}$ with a frequency $\omega_\mathbf{k}$ and polarization direction $\hat{\epsilon}_\mathbf{k}$, $\epsilon_0$ is the exciton site energy, $\{\hat{q}_n\}$ and $\{\hat{p}_n\}$ are the position and momentum operators of a set of phonon modes with frequency $\omega_p = \sqrt{K/m}$ and $m$, $\gamma$ and $\alpha$ characterizes local (Holstein) and non-local (Peierls) exciton–phonon coupling strengths, $g_c$ is the exciton-photon coupling strength and $\tau$ is a hopping parameter. As in past work on halide perovskites[28–30], we treat the lattice classically. While this model can be studied on a two-dimensional lattice, we do not expect qualitative differences from the one-dimensional case we treat here.

For a two-dimensional cavity[26] with a confinement in the $y$ direction with a length $L_y$, the transverse primary photon mode wavevector is $k_y = \pi/L_y$. Here the refractive index of the medium is assumed as 1 for simplicity. Assuming periodic boundary conditions in the $x$ direction, such that $X_{N+1} \equiv X_1$, the cavity mode wavevectors in the $x$ direction become $k_x = 0, \pm \frac{2\pi}{N \cdot L} \ldots, \pm k_{\max}$, where we choose $k_{\max} = N_c \frac{2\pi}{N \cdot L}$ as a numerical cutoff to only include energetically relevant cavity modes. Thus, the set of cavity modes are $\mathbf{k} = \{(k_x, k_y)\} = \{(0, k_y), (\pm \frac{2\pi}{N \cdot L}, k_y) \ldots, (\pm k_{\max}, k_y)\}$ with the corresponding photon frequencies $\omega_\mathbf{k} = c|\mathbf{k}| = c\sqrt{k_x^2 + k_y^2}$.

In this work, the light-matter Hamiltonian is represented in the single excited subspace $\{|e_n, 0\rangle, |g, 1_\mathbf{k}\rangle\}$, where $|e_n, 0\rangle$ represents an excitation at site $n$ and 0 photons in the cavity, and $|g, 1_\mathbf{k}\rangle$ represents a photonic excitation on the $\mathbf{k}^{\text{th}}$ mode with no excitation in the matter-subsystem. The light-matter Hamiltonian in the $\{|e_n, 0\rangle, |g, 1_\mathbf{k}\rangle\}$ basis is given as

$$\hat{H} = \sum_n (\epsilon_0 + \gamma \hat{q}_n) |e_n, 0\rangle\langle e_n, 0| + \sum_n \frac{\hat{p}_n^2}{2m} + \frac{1}{2} K \hat{q}_n^2 + \sum_\mathbf{k} \omega_\mathbf{k} |g, 1_\mathbf{k}\rangle\langle g, 1_\mathbf{k}|$$

$$+ \sum_n [\alpha(\hat{q}_{n+1} - \hat{q}_n) - \tau](|e_{n+1}, 0\rangle\langle e_n, 0| + |e_n, 0\rangle\langle e_{n+1}, 0|)$$

$$+ \sum_{\mathbf{k},n} \frac{g_c}{\sqrt{N}} \sqrt{\frac{\omega_\mathbf{k}}{\omega_0}} (\hat{x} \cdot \hat{\epsilon}_\mathbf{k}) (|e_n, 0\rangle\langle g, 1_\mathbf{k}| e^{i\mathbf{k}\cdot X_n} + |g, 1_\mathbf{k}\rangle\langle e_n, 0| e^{-i\mathbf{k}\cdot X_n}) \quad (S2)$$

We emphasize that the exciton–phonon coupling only appears in the $\{|e_n, 0\rangle\}$ subspace. This is why higher photonic character in a polariton state leads to lower effective polariton–phonon coupling. Below we have tabulated the parameters used in this work which are adapted from Ref [24,31] and not to be taken as necessarily realistic for perovskite materials.

| $\tau$ | $\alpha$ | $\gamma$ | $m$ | $K$ | $L$ |
|---|---|---|---|---|---|
| 150 cm$^{-1}$ | 1500 cm$^{-1}$/Å | 3000 cm$^{-1}$/Å | 150 amu | 14500 amu/ps$^2$ | 100 Å |

To simulate multiple photonic bands coupled to excitonic system at various detuning, we construct two different model each of which considers one exciton band and one photonic band (as was done in Fig.S4) at various detuning $\Delta_c = \epsilon_0 - \omega_0$ (with $\omega_0 = 1.57$ eV). The parameters that describe these two models (labelled as I and II) are provided below,

|  | $N$ | $N_c$ | $\Delta_c$ | $g_c$ |
|---|---|---|---|---|
| Model I | 6001 | 150 | 0.57 eV | 0.15 eV |
| Model II | 6001 | 150 | 0.2 eV | 0.15 eV |

Theoretical Polariton Bands for Model Exciton-Polariton Hamiltonian

We can obtain polariton bands for the model Hamiltonian presented in the previous section analytically if we ignore the phonons. The bands are then obtained by solving the following 2×2 Hamiltonian[32,33] (setting $\cos\theta_{\mathbf{k}} = \hat{x} \cdot \hat{\epsilon}_{\mathbf{k}}$),

$$\widehat{H}_k^0 = \begin{bmatrix} \epsilon_k & g_c\sqrt{\frac{\omega_k}{\omega_0}}\cos\theta_k \\ g_c\sqrt{\frac{\omega_k}{\omega_0}}\cos\theta_k & \omega_k \end{bmatrix}.$$

Here, $k \equiv k_x$ is used to label wavevectors instead of $\mathbf{k}$ for simplicity. The upper and lower polariton bands are then given as

$$E_\pm(k) = \frac{1}{2}(\epsilon_k + \omega_k) \pm \sqrt{(\epsilon_k - \omega_k)^2 + 4g_c^2 \cos^2\theta_k \, \omega_k/\omega_0}.$$

The Rabi-splitting $\Omega_c$ when $\omega_k = \epsilon_k$ is $\Omega_c = 2g_c\sqrt{\cos\theta_k}$. In the model considered here $\epsilon_k = \epsilon_k - 2\tau \cdot \cos(k \cdot L)$. Further, $\epsilon_k$ can be replaced with $\epsilon_k \approx \epsilon_{k=0} = \epsilon_0 - 2\tau = E_{ex}$. Note that $E_{ex} \neq \epsilon_0$.

For the parameters used in Model I and II, $E_-(k)$ roughly corresponds to the LP2 and LP3 in Fig.S4, respectively. We also emphasize that while for 2D materials in a cavity one must use an (N+1)×(N+1) Hamiltonian (where N is the number of cavity modes), it is more appropriate to use the above 2×2 Hamiltonian (1 excitonic band couples to 1 photon band) for each photonic band for a 3D material that extends along the cavity quantization direction. This is because for a 3D material, there are N degenerate exciton bands along $k_x$ (instead of 1 band for a single layer) where each of them only couples to one cavity mode with matching symmetry. That said, both models will provide the same physics showing ballistic or diffusive transport of polaritons depending on the exciton content.

## 4.2 Quantum Dynamics Approach

In this work we employ a mixed quantum-classical approach to propagate the quantum dynamics of the light-matter system. The phonon degrees of freedom (DOF) are treated classically while the electron-photon subsystem is treated quantum mechanically. The coupled motion of electronic and nuclear DOF are evolved using the mean-field Ehrenfest method.

### 4.2-a. Mean Field Ehrenfest Method

We treat the phonon DOF as nuclear degrees of freedom and evolve them classically. The photonic and excitonic DOF are evolved by solving the Time-Depended Schrödinger Equation (TDSE) for the "electronic" part (includes electronic and photonic DOF) of the Hamiltonian

$$\hat{H}_{\text{el-ph}} = \hat{H} - \sum_n \frac{\hat{p}_n^2}{2m}.$$

More precisely, the electronic-photonic wavefunction $|\Psi(t)\rangle$ is defined as

$$|\Psi(t)\rangle = \sum_j a_j(t) |\phi_j\rangle,$$

where $|\phi_j\rangle$ are the electronic-photonic states $\{|e_n, 0\rangle, |g, 1_\mathbf{k}\rangle\}$. Thus, during each nuclear step $\Delta t$ the electronic-photonic wavefunction is evolved by solving the following Time-dependent Schrodinger equation

$$i\hbar \frac{da_j(t)}{dt} = \sum_i a_i(t) \langle \phi_j | \hat{H}_{\text{el-ph}} | \phi_i \rangle$$

which is solved using RK4. The nuclear DOF in this approach is evolved with nuclear force

$$F_n(t) = \sum_{ij} a_i^*(t) a_j(t) \langle \phi_i | \nabla_n \hat{H}_{\text{el-ph}} | \phi_j \rangle.$$

In this work we have used a nuclear time-step of $\Delta t = 10$ fs to obtain converged dynamics.

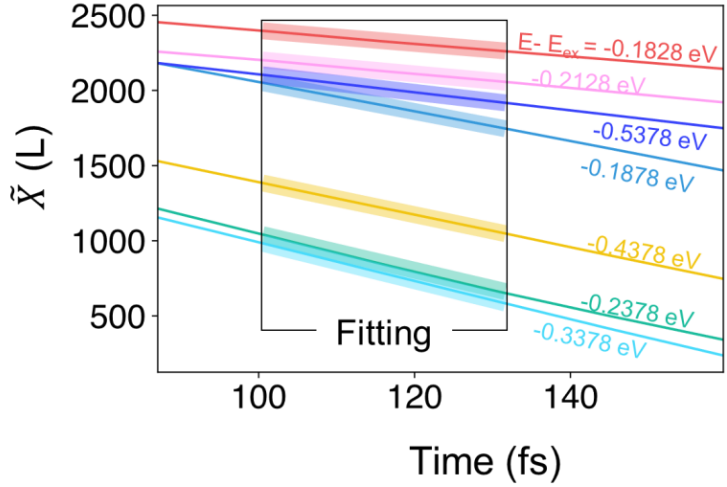

**Figure S12. Calculating the group velocity from direct numerical simulations.** Time-dependent wavefront position of the polaritonic wavepacket (solid line) initially prepared within distinct energy windows and using 250 trajectories. Transparent lines indicate linear fits of the wavefront position. The red and pink solid lines (E-E$_{ex}$ = -0.1828 eV and – 0.2128 eV) corresponds to for the model I and rest are obtained for model II.

4.2-b. Initial Conditions

We create an excitonically localized initial condition which is a linear combination of polaritonic states that lie in some specified energy window, $\left[E - \frac{\Delta E}{2}, E + \frac{\Delta E}{2}\right]$. That is

$$|\Psi(0)\rangle = \sum_j c_j(0) |\Phi_j(\mathbf{q})\rangle,$$

where $\hat{H}_{\text{el-ph}}|\Phi_j(\mathbf{q})\rangle = \mathcal{E}_j(\mathbf{q})|\Phi_j(\mathbf{q})\rangle$ and the index $j$ denotes the inclusion of states such that the condition $E - \frac{\Delta E}{2} \leq \mathcal{E}_j(\mathbf{q}) \leq E + \frac{\Delta E}{2}$ is satisfied. In this work we use $\Delta E = 0.025$ eV. The coefficients, $\{c_j(0)\}$ are obtained through a Monte Carlo procedure which maximizes excitonic localization by minimizing

$$\Delta X^2 = \sum_n \rho_n X_n^2 - \left(\sum_n \rho_n X_n\right)^2,$$

where $\rho_n = |\langle e_n, 0|\Psi(0)\rangle|^2 / \sum_m |\langle e_m, 0|\Psi(0)\rangle|^2$. Here, we also assume that the initial excitation is localized at the center, which is achieved by shifting site indexes for each trajectory such that $\langle \Psi(0)|\hat{X}|\Psi(0)\rangle = L/2$. For computing the purity of the density matrix (next section), one should also pay special attention to the relative phases between the initial state prepared in different trajectories.

The phonon DOFs are sampled from a classical Boltzmann distribution,

$$\rho(q_n, p_n) \propto e^{-\beta\left(\frac{\hat{p}_n^2}{2m} + \frac{1}{2}K\hat{q}_n^2\right)},$$

where $\beta = 1/k_B T$ (T = 300K). In this work we have used 250 trajectories to converge all results.

**4.3 Computing polariton coherence through the purity of the density matrix**

To understand how phonons lead to decoherence of polaritonic wavepackets, and to confirm that ballistic transport is associated with polariton coherence, we compute the time-dependent purity of the reduced density matrix $\hat{\rho}(t) = |\Psi(t)\rangle\langle\Psi(t)|$ of the exciton-photon subsystem. The purity is defined as the trace of the density matrix squared, $Tr[\hat{\rho}^2(t)]$. Note that the initial localized wavepacket prepared depends on the phonon coordinates $\{q_n\}$ which assume different values for different trajectories; therefore, the purity at $t = 0$ is less than 1. We also disregard trajectories that have less than 50% overlap to some reference trajectory. Note that for low exciton content (<25% excitonic) the initial purity is very close to 1. To compare the decay of the purity at various excitation energies, we present the normalized purity $Tr[\hat{\rho}^2(t)]/Tr[\hat{\rho}^2(0)]$, shown in Figure S13 for different excitonic fractions of the polariton. The normalized purity is 1 for a highly coherent system, and decays to 0 in the limit of an incoherent system. As expected, these results show that the decay of the purity increases with increasing exciton content. For low exciton content, such as EPs I (6.5% excitonic), II (9.0%), and III (15%), the decay of the purity is negligible over the EP lifetime. For higher exciton content (IV 25%), some decay of purity occurs on timescale of several hundreds of fs, indicating coherence is still maintained for much of the polariton lifetime. At 47% exciton content (V), the purity decays to 0.1 within 300 fs; correspondingly, we do not observe a ballistic wavefront in our simulations for 47% exciton content. These calculations confirm that the observation of ballistic propagation is directly associated with coherence of the polariton subsystem.

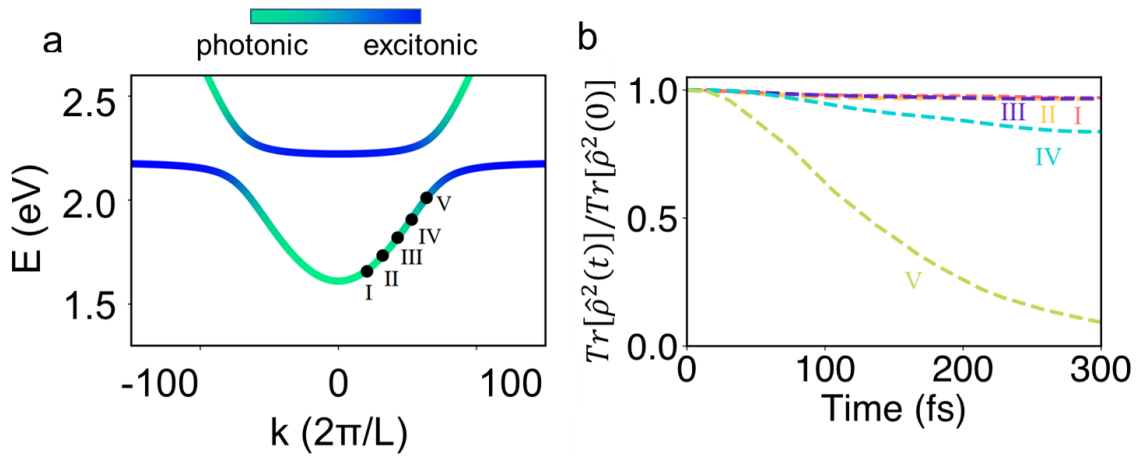

**Fig. S13. Computing purity.** (a) Polaritonic bands with color coding that represent the excitation character. (b) Time dependent decay of the purity for initial wavepackets prepared at various energy windows, labeled as I−V as highlighted in (a). The exciton content of polaritons I−V are 6.5 %, 9.0 %, 15 %, 25 % and 47 %, respectively.

## 4.4 Numerical Details

To extract the velocities of wavefronts from our simulations, we compute the position of the wavefront $\tilde{X}(t) = L \cdot N'$ where $N'$ is determined from is obtained by solving from the expression

$$\sum_{n=0}^{N'} \rho_n = P_X$$

where $\rho_n = |\langle -, n|\Psi(t)\rangle|^2 / \sum_m |\langle -, n|\Psi(t)\rangle|^2$, and $P_X = 0.04$ is chosen to define the position of the wavefront such that 4% of excitonic probability has crossed $\tilde{X}(t)$ at time $t$. Note that $|-,n\rangle = \frac{1}{\sqrt{N}} \sum_k e^{ikR_n} |-,k\rangle$. The velocity of the wavefront is then computed by fitting $\tilde{X}(t)$.

Figure S14 presents the time-dependent wavefront position of the polaritonic wavepacket initially prepared within various energy windows. We obtain group velocities by linearly fitting between 72-188 fs. These group velocities are reported in the main text. These simulations were done using the Extreme Science and Engineering Discovery Environment (XSEDE)[34] and specifically Open Science Grid Consortium.[35,36]

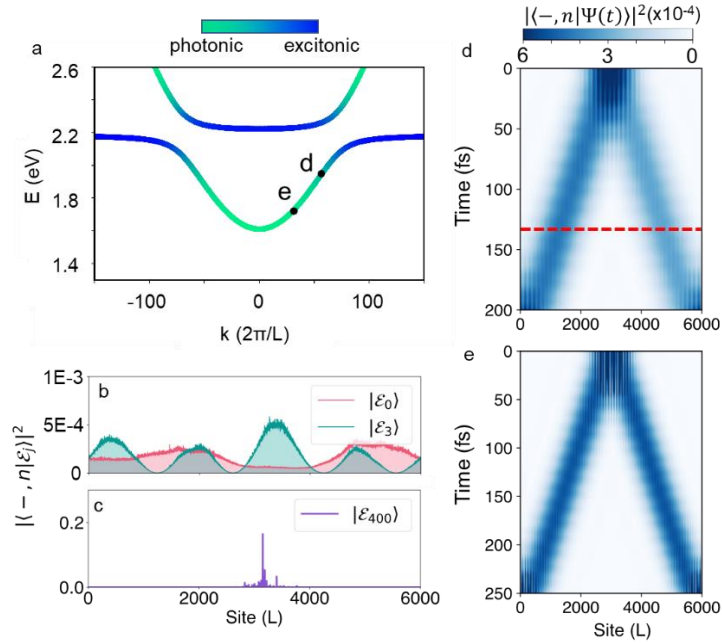

**Figure S14. Quantum dynamical simulation of exciton-polariton motion.** (a) Polaritonic bands (ignoring phonons) with color coding that represent the excitation character. (b) Polaritonic eigenstates $|\mathcal{E}_0\rangle$ and $|\mathcal{E}_3\rangle$ (6.6% excitonic character) that are delocalized in site (position) space. (c) Polaritonic eigenstate $|\mathcal{E}_{400}\rangle$ that lies around $\approx 2.11$ eV (98.9% excitonic) which is localized in site space. (d) Ballistic spreading of polaritonic probability density for the same detuning as in Figure 2d of the main text, but in the absence of exciton–phonon coupling (setting $\gamma = \alpha = 0$). The wave front of the polariton wave-packet propagates always ahead of the one with phonon scattering. The line cut at 130 fs is plotted in figure 2e of the main text. (e) Ballistic spreading of the polaritonic probability density for the excitation labeled 'e' in panel (a), without phonon interactions.

# Supplementary Note 5

## 5.1 Additional data and fits for EP propagation at room temperature under non-resonant excitation

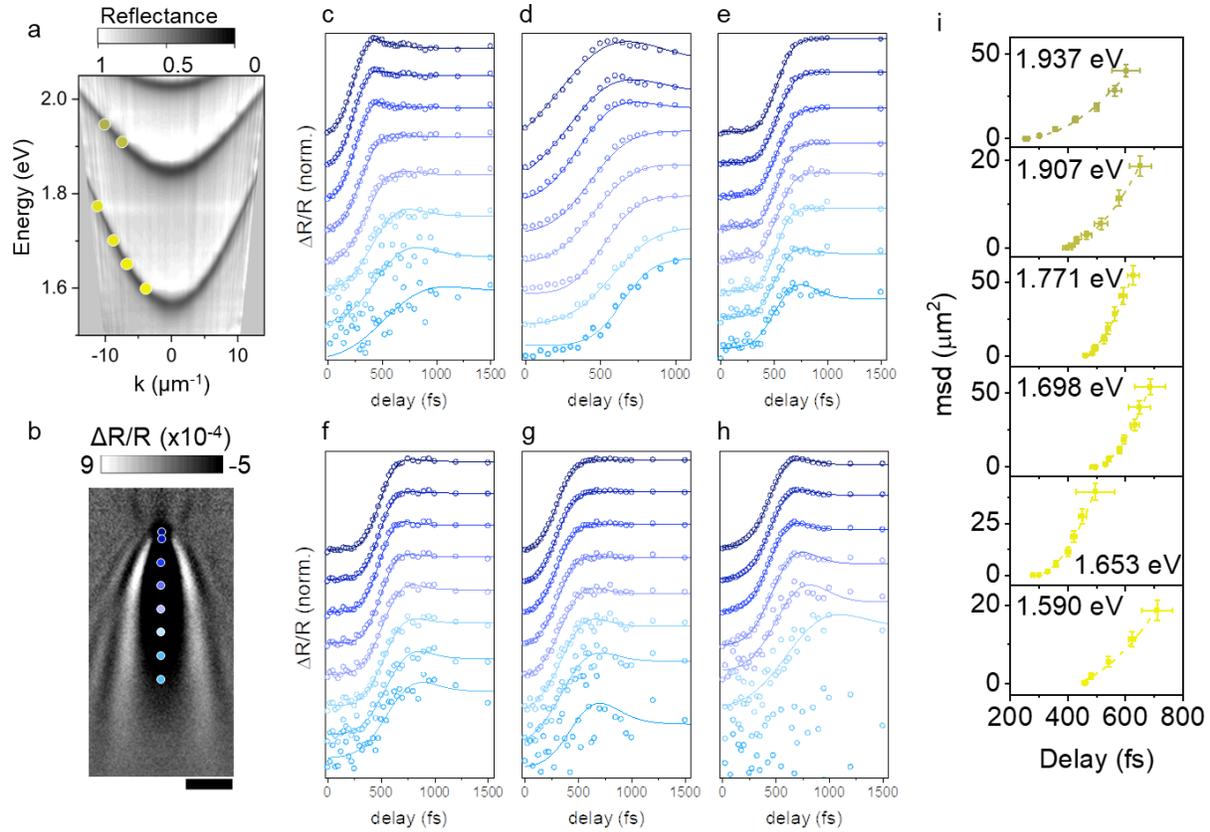

**Figure S15. Fitting of propagation data for sample 1 in Figure 2 of the main text**. (a) Angle-resolved reflectance spectrum. Probe energies and momenta used to probe different EPs are indicated with filled circles, color-coded with panel (i). (b) Blue dots represent spatial locations for which we fit the signal grow-in time to extract the polariton propagation velocity. From bottom to top, the distances are 0, 0.32, 1.33, 2.35, 3.37, 4.32, 5.33, and 6.35 μm away from excitation location. Scale bar is 2 μm. (c-h) EP transport analysis for each EP. From (c) to (h), the EP energies are 1.94 eV, 1.91 eV, 1.77 eV, 1.70 eV, 1.65 eV and 1.59 eV, respectively (exciton content are 31.5 %, 25.9 %, 12.2 %, 8.8 %, 7.4 %, and 5.9 %). The 1.59 eV EP showed relatively short propagation, so the last two position do not display any grow-in. (i) msd extracted from panels c-h. Dashed lines correspond to ballistic propagation fits. The resulting velocity is plotted in Figure 2 of the main text. Error bars are one standard deviation

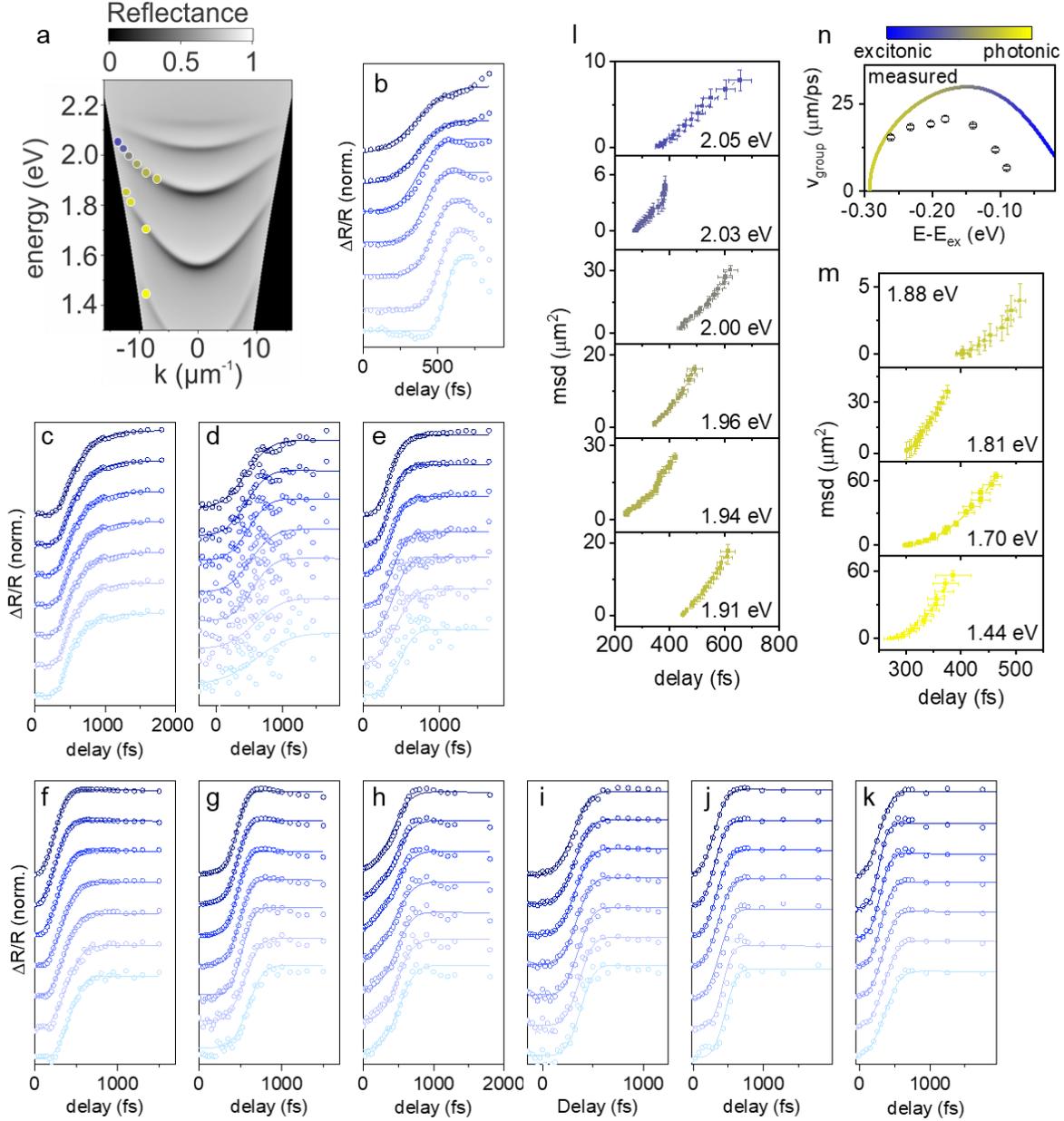

**Figure S16. Fitting of propagation data for sample 2 in Figure 2 of the main text.** (a) Angle-resolved reflectance spectrum. Probe energies and momenta used to probe different EPs are indicated with filled circles, color-coded with panels (l) and (m). (b-k) EP transport analysis for each EP . From (b) to (k), the EP energies are 2.05 eV, 2.03 eV, 2.00 eV, 1.96 eV, 1.94 eV, 1.91 eV, 1.88 eV, 1.81 eV, 1.70 eV and 1.44 eV respectively (exciton content are 69.7 %, 62.1 %, 49.0 %, 36.5 %, 31.5 %, 25.9 %, 21.7 %, 14.4 %, 8.8 %, and 3.8 %). (l,m) EP msd for each dataset. All EPs below 2.00 eV use ballistic transport fits. EPs with energies of 2.05 eV (exciton content 69.7 %) and 2.03 eV (exciton content 62.1 %) lose coherence with phonon scattering, and are fitted with a diffusive transport model (msd $\propto t$). (n) Expected group velocity (solid line) *vs.* measured transport velocity (symbols) for each probing condition, showing an increasing deviation as exciton content is increased. Error bars are one standard deviation

## 5.2 EP propagation at 5 K under non-resonant excitation

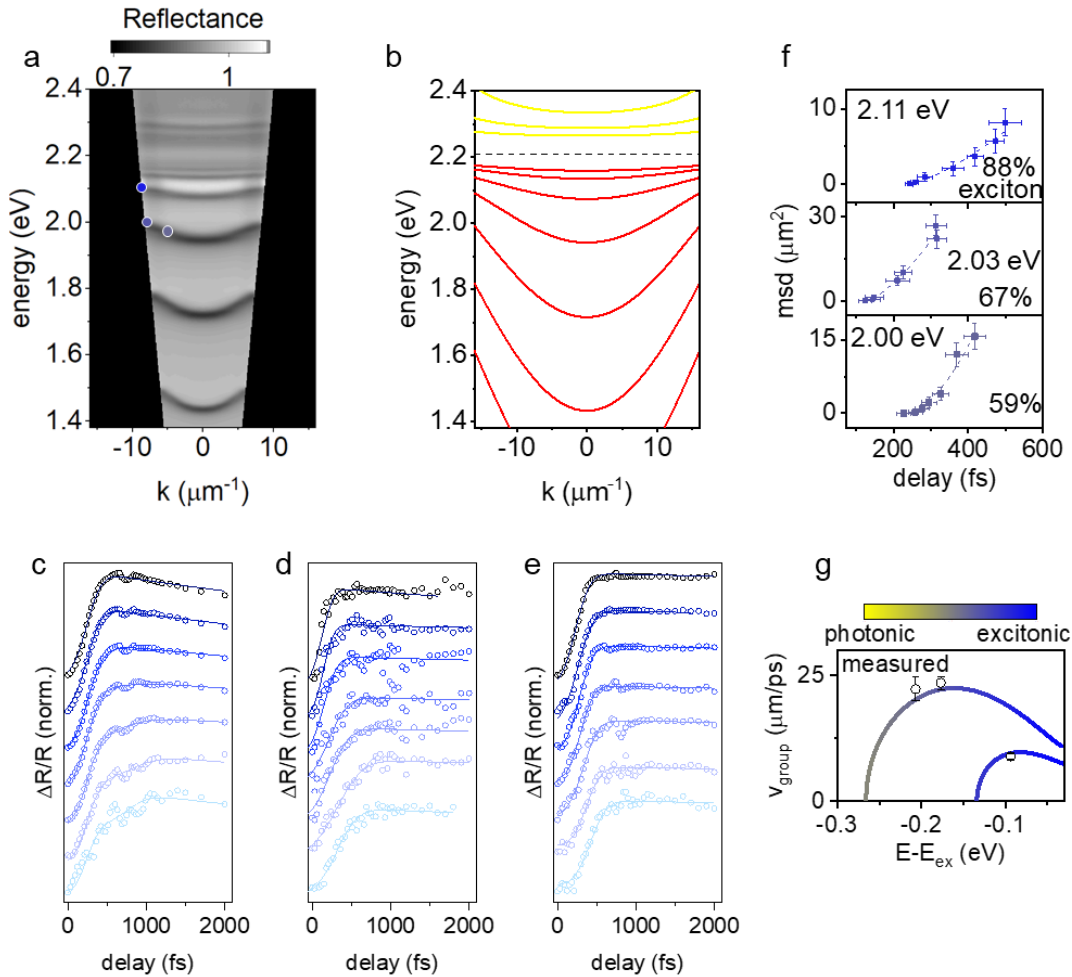

**Figure S17. Fitting of propagation data at cryogenic temperatures (sample 3 in Figure 2 of the main text).** (a) Angle-resolved reflectance spectrum obtained at 5 K in the cryostat. Note the lower range of momenta available due to the lower-NA objective (0.9 NA) compared to other datasets. The narrower exciton linewidth, and reduced absorption between the optical gap and the electronic bandgap, allows resolving the upper polariton branches in this sample. (b) Coupled oscillator model fit to the data in panel (a). The much larger oscillator strength of the exciton transition at 5 K increases the Rabi splitting to 500 meV. (c-e) EP transport analysis. From (c) to (e), the EP energies are 2.11 eV, 2.03 eV, 2.00 eV (exciton content are 87.7%, 66.6 %, and 59.4%). (f) EP msd for each dataset, with a ballistic transport fit. (g) Expected group velocity (solid line) *vs.* measured transport velocity (symbols) for each probing condition. The measured velocities match the experimental dispersion within experimental error. Note that exciton transport is not affected by the phase transition of $BA_2(MA)Pb_2I_7$ at 270 K[37], making this perovskite a good model system for temperature-dependent EP studies. Error bars are one standard deviation

## 5.3 Spatial coherence through double-slit interference

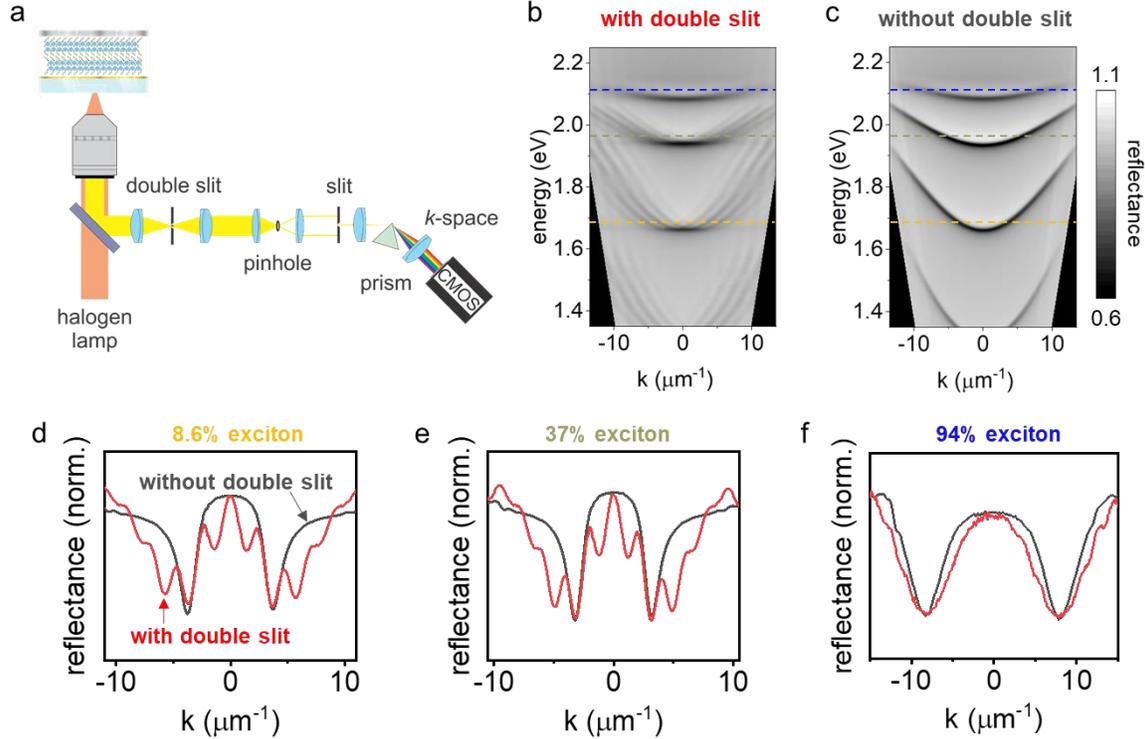

**Figure S18. Establishing spatial coherence through double-slit interference measurements at room temperature**.[38–40] (a) Optical setup used to measure spatial coherence. An incoherent halogen white light source is used, and a double slit is inserted in the first image plane. The slits are separated by 0.3 mm, corresponding to 4.76 microns on the sample image plane after accounting for the system magnification. (b) Angle-resolved reflectance spectrum with the double slit inserted, exhibiting clear replicas of the lower polariton branches. (c) Corresponding spectrum obtained without the double slit. (d-f) Line cuts along $k$ for three different energies showing clearly-resolved interference fringes when the double slit is inserted. The fringes are not visible for the highest-energy branch, suggesting lower spatial coherence for high-exciton EPs. Note that linewidth broadening at high exciton content also contributes to lower fringe visibility. Although these double-slit measurements confirm that EPs exhibit spatial coherence in the system, they do not report on any dynamic (e.g. scattering) effects that contribute to incoherent propagation.

## 5.4 EP propagation at room temperature under resonant excitation

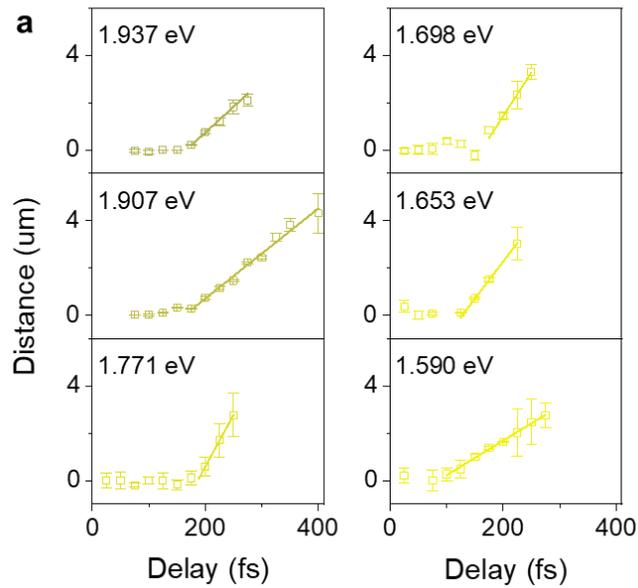

**Figure S19. EP propagation under resonant excitation.** (a) Exciton-polariton transport analysis in sample 1 under resonant excitation conditions (complementary data to Figure 3f). The initial apparently immobile signal is assigned to EP-EP scattering following resonant excitation, which creates a dense population of EPs within the pump pulse temporal width of ~60 fs. In Figure 3f of the main text, we show Monte-Carlo simulations that reproduce this behavior.

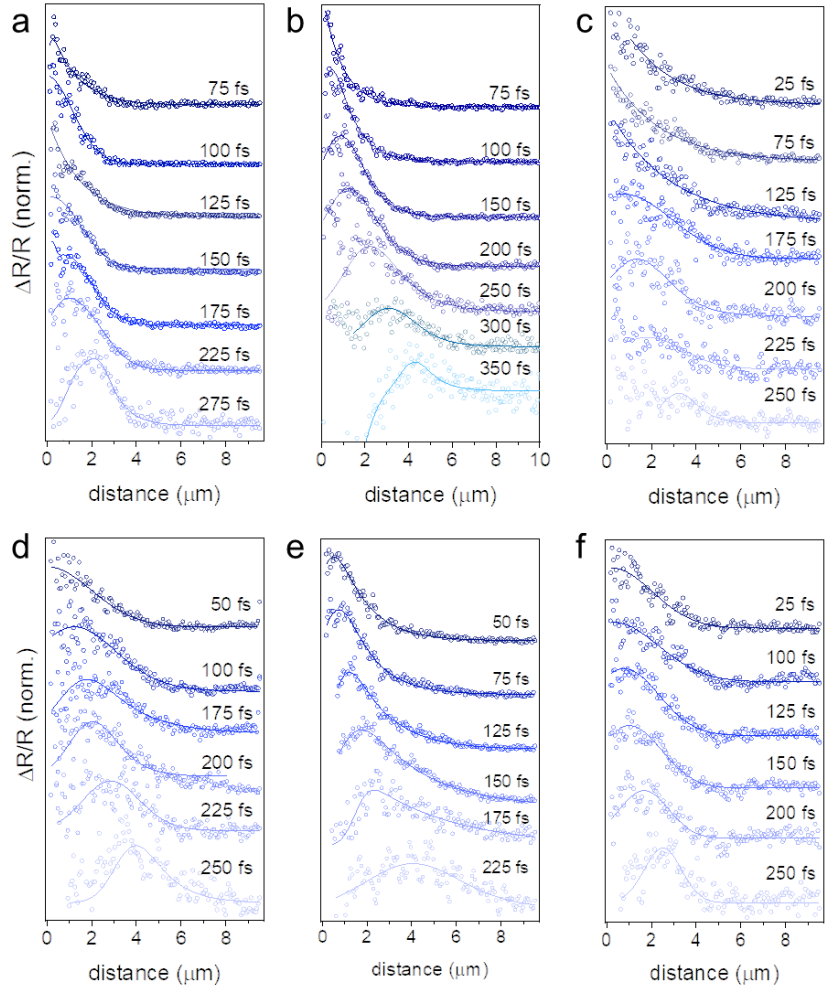

**Figure S20. Data fits for the resonant excitation EP propagation data presented in Figure S19**. (a-f) Temporal evolution series for EP energies of 1.94 eV, 1.91 eV, 1.77 eV, 1.70 eV, 1.65 eV and 1.59 eV, respectively. Note that for resonant excitation, a clear wavepacket-like feature is observed because there are no reservoir exciton states refilling the LP branch and causing the elongated features observed in non-resonant excitation. As such, fitting for resonant excitation experiments is performed by tracking the position of the maximum amplitude of the wavepacket, as illustrated in the figure.

## 5.5 Characterization of cavity-free polaritonic slabs

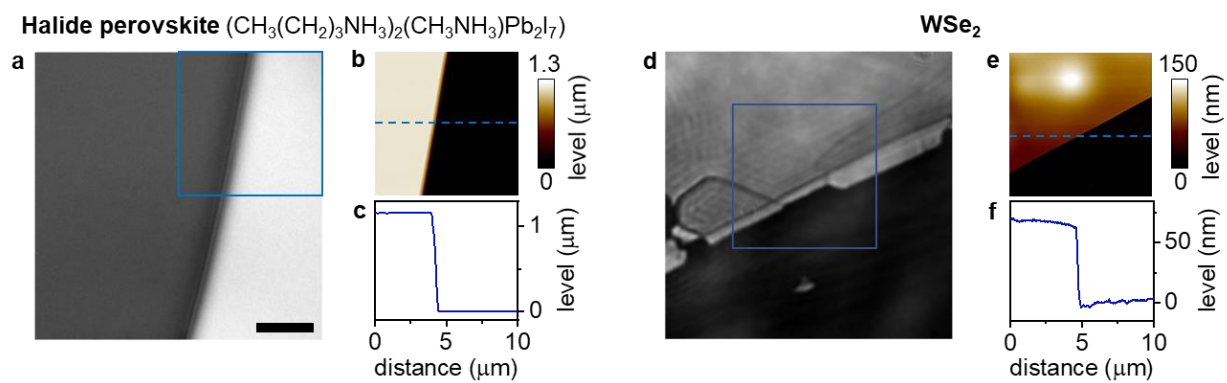

Figure S21. Optical images (a,d), atomic force microscopy images (b, e) and corresponding line cuts (c, f) for the layered halide perovskite and $WSe_2$ slabs used for Figures 4b and 4c. The layered perovskite slab is 1.13 μm thick. The $WSe_2$ slab is 69 nm thick.